\documentclass[prd,tightenlines,twoside,secnumarabic,onecolumn,floatfix,nofootinbib,showpacs,11pt]{revtex4-1}

\pdfoutput=1
\usepackage[english]{babel}
\usepackage{calligra}
\usepackage{epsfig}
\usepackage{float}
\usepackage{multirow}
\usepackage{epstopdf}
\usepackage{amsmath,amssymb,bm,slashed}
\usepackage{mathtools}
\usepackage{graphicx}
\usepackage[sort&compress]{natbib}
\usepackage{xcolor}
\usepackage[normalem]{ulem}
\usepackage{hyperref}
\usepackage{cleveref}
\usepackage{subfigure} 
\definecolor{red}{rgb}{1.0, 0, 0}
\usepackage{slashed}
\usepackage{ctable}

\allowdisplaybreaks

\bibpunct{[}{]}{,}{n}{}{,}

\newcommand{\ev}[1]{\ensuremath{\left\langle #1 %
                     \right\rangle}} 

\newcommand{\BR}{\text{BR}}

\newcommand{\parenbar}[1]{\overset{
            \raisebox{-0.15em}{\scalebox{.4}{\textbf{(}}}
            \raisebox{-0.3em}{{\hspace{.03em}--\hspace{.05em}}}
            \raisebox{-0.15em}{\scalebox{.4}{\textbf{)}}}} {#1}}

\renewcommand{\vec}[1]{{\mathbf{#1}}}

\begin{document}

\title{Boosted Dark Matter in IceCube and at the Galactic Center}
\author{Joachim Kopp}        \email[Email: ]{jkopp@uni-mainz.de}
\author{Jia Liu}             \email[Email: ]{liuj@uni-mainz.de}
\author{Xiao-Ping Wang}      \email[Email: ]{xiaowang@uni-mainz.de}
\affiliation{PRISMA Cluster of Excellence and Mainz Institute for Theoretical Physics,
Johannes Gutenberg University, 55099 Mainz, Germany}
\date{\today} 

\begin{abstract}
We show that the event excess observed by the IceCube collaboration at TeV--PeV
energies, usually interpreted as evidence for astrophysical neutrinos, can be
explained alternatively by the scattering of highly boosted dark matter
particles. Specifically, we consider a scenario where a $\sim 4$~PeV scalar
dark matter particle $\phi$ can decay to a much lighter dark fermion $\chi$,
which in turn scatters off nuclei in the IceCube detector. Besides these
events, which are exclusively shower-like, the model also predicts a secondary
population of events at $\mathcal{O}(100 \text{TeV})$ originating from the
3-body decay $\phi \to \chi \bar\chi a$, where $a$ is a pseudoscalar which
mediates dark matter--Standard Model interactions and whose decay products
include neutrinos.  This secondary population also includes track-like events,
and both populations together provide an excellent fit to the IceCube data.  We
then argue that a relic abundance of light Dark Matter particles $\chi$, which
may constitute a subdominant component of the Dark Matter in the Universe, can
have exactly the right properties to explain the observed excess in GeV gamma
rays from the galactic center region.  Our boosted Dark Matter scenario also
predicts fluxes of $\mathcal{O}(10)$~TeV positrons and $\mathcal{O}(100
\text{TeV})$ photons from 3-body cascade decays of the heavy Dark Matter
particle $\phi$, and we show how these can be used to constrain parts of the
viable parameter space of the model.  Direct detection limits are weak due to
the pseudoscalar couplings of $\chi$. Accelerator constraints on the
pseudoscalar mediator $a$ lead to the conclusion that the preferred mass of $a$
is $\gtrsim 10$~GeV and that large coupling to $b$ quarks but suppressed or vanishing
coupling to leptons are preferred.
\end{abstract}

\begin{flushright}
  MITP/15-014
\end{flushright}

\maketitle

\section{Introduction}
\label{sec:intro}

The IceCube experiment at the South Pole has recently made international
headlines by discovering an excess of events in the energy range from 30~TeV to
2~PeV~\cite{Aartsen:2013jdh,Kappes:2015woa,Aartsen:2014gkd}. These events are
usually interpreted as evidence for a flux of astrophysical neutrinos with a
power-law spectrum $\sim E_\nu^{-2}$, originating from the production and
subsequent decay of charged pions, kaons, muons and neutrons produced in
collisions of ultra-high energy charged cosmic rays with protons or photons in
astrophysical sources.

Despite the plausibility of this explanation, there are other possibilities.
For instance, ultra-high energy neutrinos could be produced in non-standard
processes such as the decay~\cite{Feldstein:2013kka, Esmaili:2013gha,
Bai:2013nga, Higaki:2014dwa, Rott:2014kfa, Esmaili:2014rma, Fong:2014bsa,
Dudas:2014bca} or
annihilation~\cite{Zavala:2014dla, Chen:2014lla} of very heavy DM particles, or
in the early decay of ultra-massive long lived particles~\cite{Ema:2013nda,
Ema:2014ufa}.

In this paper, we explore another alternative idea, namely that IceCube may be
observing dark matter (DM)
particles with PeV energy \emph{directly} (as opposed to observing only
neutrinos from their annihilation or decay). The idea, which has first
been put forward in~\cite{Bhattacharya:2014yha}, is the following: a heavy
$\mathcal{O}(\text{PeV})$ DM species $\phi$, which makes up a substantial
fraction of the dark matter in the Universe, decays to a much lighter species
$\chi$. The resulting flux of highly boosted $\chi$ particles scatters on
nuclei in the IceCube detector and leads to the observed energy deposits
$E_\text{dep}$ up to few PeV. An upper cutoff on $E_\text{dep}$ is naturally
provided by the mass of $\phi$, explaining the absence of events above a few PeV.

The idea of direct detection of boosted DM in large volume terrestrial
experiments was introduced in ref.~\cite{Agashe:2014yua} in the context of new
light ($\mathcal{O}(1)$~GeV) particles produced in the annihilation of
$\mathcal{O}(100)$~GeV DM particles in the galactic halo.  The authors focused on
electron recoil signatures in Super-Kamiokande~\cite{Fukuda:2002uc},
Hyper-Kamiokande~\cite{Abe:2011ts} and PINGU~\cite{Aartsen:2014oha}.  In a
subsequent paper~\cite{Berger:2014sqa}, also the possibility of detecting
boosted particles from the annihilation of heavy DM captured in the Sun has
been considered. Such signals can be enhanced if the heavy DM particles are
self-interacting, so that their capture rate in the Sun is
increased~\cite{Kong:2014mia}. Also a model with ``dark nucleosynthesis''
could lead to mildly boosted dark sector particles emerging from
the Sun~\cite{Detmold:2014qqa}.
The detection of boosted DM annihilation
products at much lower energies $\lesssim \text{GeV}$ in direct DM detection
experiments is discussed in~\cite{Cherry:2015oca}. The recoil energy spectrum
and the annual modulation signal in this case are very distinct from those
expected from scattering of ordinary non-relativistic DM.
Compared to these previous works which focus on boosted DM with
energies of $\lesssim 100$~GeV, we study signals at even higher energies
up to $\mathcal{O}(\text{PeV})$, and we consider not only direct DM searches,
but also indirect signatures which may be very relevant in our model.

In the context of the conventional neutrino interpretation of the IceCube events, there is
an ongoing debate about the neutrino flavor ratios required to explain the
data.  The generic expectation for neutrino production from pion decay is that
the flavor composition of the astrophysical flux at the source (S) is $(\nu_e
:\nu_\mu : \nu_\tau)_S = (1:2:0)_S$. After propagation and oscillation, the
final flux at Earth (E) would have a composition of $(\nu_e :\nu_\mu :
\nu_\tau)_E \approx (1:1:1)_E$. The IceCube events are categorized as track
events and shower events, where the former are mostly from $\nu_\mu$ charged current
(CC) interactions with nucleons in which the produced high energy muon leaves a
track in the detector. The shower events are attributed either to neutral
current (NC) interactions of neutrinos or to charged current interactions of
$\nu_e$ and $\nu_\tau$. In several analyses, the flavor ratios of
the IceCube events have been studied~\cite{Chen:2013dza, Mena:2014sja,
Palomares-Ruiz:2014zra, Watanabe:2014qua, Palomares-Ruiz:2015mka, Palladino:2015zua, Aartsen:2015ivb}, and while in general, the data appears
consistent with a $(1:1:1)_E$ flavor ratio, a mild lack of $\nu_\mu$
has been found.

A unique feature of the boosted DM scenario is that only shower events are
predicted at PeV energy, while at the lower energies, the ratio of track events
to shower events is similar to what is expected in the canonical interpretation
of the data in terms of astrophysical neutrinos.  The reason our model
predicts also track events at low energy is that, in addition to the dominant
flux of boosted $\chi$ particles, also a secondary flux of DM-induced neutrinos
is expected.  It arises when the particle that mediates DM--SM
interactions---taken to be a pseudoscalar $a$ here---is directly produced as final state
radiation in the heavy DM decay, $\phi \to \chi \bar\chi a$, and subsequently
decays to SM particles.  While the primary contribution to the IceCube data
from $\chi$ scattering peaks at PeV energies but drops at lower energies due
to the properties of the pseudoscalar interaction, the secondary neutrino
flux peaks at $\mathcal{O}(100\ \text{TeV})$ energies.  Thus, our scenario
is also able to explain not only the observed ratio of shower to track events,
but also the mild (though not yet significant) deficit of
events in the intermediate energy range of $\text{few} \times 100$~TeV.

Note that the IceCube collaboration has recently published a new
analysis~\cite{Aartsen:2014muf}, the results of which are given
separately for events coming from above, i.e.\ from the southern sky,
and from below, i.e.\ from the northern sky.
This analysis features a notable, but not yet
statistically significant, bump in the event spectrum from the southern sky at
around $80$~TeV.  Since the galactic center is located in the southern
hemisphere, a decaying DM scenario like ours predicts a larger contribution
from the southern sky than from the northern sky.  Thus, this bump could be
potentially interpreted as being due to the secondary neutrino flux discussed
in the previous paragraph, which peaks at around $100$~TeV.

In addition to the new window to the high energy Universe opened by IceCube,
also observations at lower energies $\sim$~GeV have caused a stir recently.
Namely, an excess of gamma rays from the vicinity of the galactic center was
found in Fermi-LAT data, which could be explained by DM annihilation
\cite{Goodenough:2009gk, Hooper:2010mq, Daylan:2014rsa}.  A good fit to the
Fermi-LAT data is obtained for instance for a 30--40~GeV DM particle
annihilating to $b \bar{b}$ with a thermally averaged cross-section of about
$\ev{\sigma v_\text{rel}} \sim 10^{-26}\ \text{cm$^3$ sec$^{-1}$}$, similar to
the annihilation cross section expected for a thermal relic. In our scenario, a
subdominant primordial population of the light DM species $\chi$ can naturally
provide such signal by annihilation through $s$-channel exchange of the
pseudoscalar mediator $a$. We will demonstrate that there is a viable region of
parameter space which can explain the Fermi-LAT gamma ray signal and the IceCube signal
simultaneously.

In the following, we first introduce our toy model of boosted DM
in sec.~\ref{sec:models} and then discuss the expected IceCube signals in
sec.~\ref{sec:ic}.  In particular, we show which regions of parameter space
could explain the recently observed high-energy events. In
sec.~\ref{sec:relic-density} we review mechanisms for explaining the observed
DM relic density~\cite{Ade:2013zuv} in the boosted DM scenario, and in sec.~\ref{sec:GCE} we
discuss the possibility that the galactic center gamma ray excess is
explained by $\chi \bar \chi$ annihilation along with the IceCube PeV events.
We then discuss other constraints on the model in sec.~\ref{sec:constraints},
in particular limits from measurements of the cosmic positron and
electron spectrum~\cite{Aguilar:2013qda, Accardo:2014lma, Aguilar:2014mma,
Ackermann:2010ij, Aharonian:2008aa,Aharonian:2009ah}, from isotropic diffuse
gamma rays~\cite{Ackermann:2014usa, ::2014sla}, from direct detection
experiments and from searches for the pseudoscalar mediator $a$ in flavor
physics experiments and at high energy colliders.  We summarize and conclude in
sec.~\ref{sec:conclusions}.

\section{The framework}
\label{sec:models}

While most of the qualitative results of this paper apply to any PeV-scale
boosted DM model, we consider as a specific example a toy model
featuring a dark sector that contains two DM particles: a heavy real scalar
$\phi$ with mass $m_\phi \sim \mathcal{O}(\text{PeV})$ and a light Dirac
fermion $\chi$ with mass $m_\chi \sim \mathcal{O}(10)$~GeV.  We denote
the relic abundance of $\phi$ by $f_\phi \Omega_\text{DM}$ and the
relic abundance of $\chi$ by $f_\chi \Omega_\text{DM}$, where
$\Omega_\text{DM} \simeq 0.258$ is the total dark matter density
in the Universe~\cite{Planck:2015xua}.  We will discuss in
sec.~\ref{sec:relic-density} how $f_\phi$ and $f_\chi$ could be determined
in the early Universe.  We assume that there are no other dark relics
besides $\phi$ and $\chi$, i.e.\ we assume $f_\phi + f_\chi = 1$.
The dark sector Lagrangian reads
\begin{align}
  \mathcal{L}_\text{DS} \equiv \frac{1}{2} (\partial^\mu \phi) (\partial_\mu \phi)
                             - \frac{1}{2} m_\phi^2 \phi^2
                             + i \bar{\chi} \slashed{\partial} \chi
                             - m_\chi \bar\chi \chi
                             - y_{\phi\chi} \phi \bar{\chi} \chi \,.
  \label{eq:lagrangian}
\end{align}
Here, the coupling constant $y_{\phi\chi}$ determines the $\phi \to \chi
\bar\chi$ decay rate. We
assume $y_{\phi\chi}$ to be tiny, so that the lifetime of $\phi$ is
significantly longer than the age of the Universe.
One possible way of explaining the smallness of $y_{\phi\chi}$ could be
to envision $\phi$ as a composite particle made up of superheavy
constituents $Q_\phi$ and held together by a new confining gauge interaction.
When this new gauge symmetry is broken by a tiny amount, a correspondingly small
mixing between the $Q_\phi$ and $\chi$ could be generated.
Note that we do not include quartic couplings of $\phi$ or Higgs portal
couplings in eq.~\eqref{eq:lagrangian} since these interactions will
not be relevant to our phenomenological discussion.  A possibly problematic
term could be an operator of the form $\phi (H^\dag H)$, but we assume
the mechanism that suppresses $y_{\phi\chi}$ also forbids or suppresses
this operator.

The light DM $\chi$ interacts with SM particles through a pseudoscalar
mediator $a$~\cite{Boehm:2014hva,Berlin:2014tja, Dolan:2014ska,Arina:2014yna}.
This pseudoscalar couples to light DM and Standard Model fermions through the
Lagrangian
\begin{align}
  \mathcal{L}_\text{int} \equiv i g_\chi a \bar\chi \gamma_5 \chi
                              + i \sum_f g_{Y_f} \frac{\sqrt{2} m_f}{v} \,
                                a \, \bar f\gamma _5 f \,,
  \label{eq:lagrangian2}
\end{align}
where $g_\chi$ and $g_{Y_f}$ are real couplings of $a$ to light DM $\chi$ and to
Standard Model fermions $f$, respectively, $m_f$ are the SM fermion masses and
$v \simeq 246\ \text{GeV}$ is the vacuum expectation value (vev) of the SM
Higgs field.  While generically all $g_{Y_f}$ are free parameters, we will
specifically consider natural scenarios in which the $g_{Y_f}$ are
generation-independent.

\begin{table}
  \centering
  \begin{ruledtabular}
  \begin{tabular}{l@{\quad}cccccc|cccl}
         & $m_a$ & $m_\phi$ & $m_\chi$ & $g_{Y_b}$ & $g_\chi$ & $\tau_\phi/f_\phi$ & $\ev{\sigma v_\text{rel}}_{b\bar{b}}$ & $f_\chi$ & $\BR_3(\phi\!\to\!\chi \bar\chi a)$ & Comment \\
         & [GeV] & [PeV]    & [GeV]    &           &          & [$10^{25}$ s]      & [$10^{-26}$ cm$^3$/s]                 &          &                                 &         \\
    \hline
    BP 1 &  12   &   4.5    &   30     &  0.86     &  0.396   & $3.6$              & $2.8$                                 &  0.6     & 0.022                          & \textit{Vector-like} model only \\
    BP 2 &  80   &   3.9    &   30     &  1.51     &  0.462   & $1.8$              & $18$                                  &  0.33     & 0.026                           &         \\
  \end{tabular}
  \end{ruledtabular}
  \caption{Summary of our two benchmark points (BP), both of which can explain
    the IceCube event excess and the galactic center gamma ray excess.  In both
    models, the pseudoscalar $a$ is assumed to couple dominantly to $b$ quarks.
    We also give the calculated values of the velocity-averaged $\chi$
    annihilation cross section $\ev{\sigma v_\text{rel}}_{b\bar{b}}$ (relevant
    for the galactic center gamma ray excess), of the fractional abundance of
    the light DM species $f_\chi = 1 - f_\phi$ and of the branching ratio for
    the radiative decay $\phi \to \chi \bar\chi a$. Note that benchmark point~1
    can be realized only in the \textit{Vector-like} quark model since in the
    \textit{MSSM-like} and \textit{Flipped} scenarios, laboratory constraints
    on $g_{Y_b}$ are too strong (see sec.~\ref{sec:constraints}).}
  \label{tab:benchmarks}
\end{table}

Throughout most of the paper, we will consider two benchmark points in the
parameter space of the model, defined in table~\ref{tab:benchmarks}.  The heavy
DM mass $m_\phi$ and lifetime $\tau_\phi$, the light DM mass $m_\chi$, and the
couplings $g_\chi$ and $g_{Y_b}$ are chosen such that both the IceCube excess
of high energy events as well as the galactic center gamma ray excess are
explained.  We assume the mass of $a$ to satisfy $m_a \gtrsim 10$~GeV since
constraints are weak in this case (see sec.~\ref{sec:constraints}), thus
allowing large couplings $g_{Y_f}$ to fermions.  This is important for the model to
fit the IceCube data and is also interesting because it allows for a detectable
indirect signal from the annihilation of non-relativistic relic $\chi$
particles.

Since the coupling of the pseudoscalar $a$ to SM fermions in
eq.~\eqref{eq:lagrangian2} should be considered as an effective operator after
the spontaneous breaking of electroweak symmetry, we need to discuss possible
ultraviolet completions for such an operator.  We consider here three
interesting models which can provide such a coupling.

{\bf MSSM-like model.} In the first model, the pseudoscalar $a$ mixes with an
extended Higgs sector, for example
with the pseudoscalar $A^0$ in a type-II Two Higgs Doublet Model (2HDM), by
a term of the form $i a H_1^\dag H_2 + h.c.$~\cite{Ipek:2014gua}.
In this case, the Higgs couplings to quarks and leptons are the same as
in the Minimal Supersymmetric Standard Model (MSSM). We therefore denote this
model as \textit{MSSM-like}. The relations for the couplings between
effective operator model and the complete renormalizable model read \cite{Ipek:2014gua}
\begin{align}
  & g_{Y_d} = g_{Y_\ell } =  -\tan{\beta} \sin{\theta} / \sqrt 2 \\
  & g_{Y_u} = -\cot{\beta} \sin{\theta} / \sqrt 2,
\end{align}
where $\tan\beta = v_2 / v_1$ is the ratio of the two Higgs vevs and $\sin{\theta}$ is the
mixing angle between the pseudoscalar $a$ and the $A^0$ boson of the 2HDM.
$g_{Y_d}$, $g_{Y_\ell}$ and $g_{Y_u}$ are the generation-independent
normalization factors of the
Yukawa-like couplings for down-type quarks, leptons and up-type quarks,
respectively. Since the pseudoscalar $a$ couples to SM fermions only through
its mixing with $A^0$, all of these couplings are suppressed by $\sin{\theta}$.
As mentioned in the Introduction, we are interested in particular in scenarios
with large coupling between the pseudoscalar $a$ and bottom quarks to
optimally fit the galactic
center gamma ray excess. This requires large $\tan{\beta}$ to lift up the
coupling to down-type quarks.  Already at this stage, we can see that the
\textit{MSSM-like} model will be constrained by experiments sensitive to
anomalous couplings of the charged leptons (which are also $\tan\beta$-enhanced)
and by searches for an extended
Higgs sector.  As we will see in sec.~\ref{sec:constraints}, these
constraints lead to the conclusion that the IceCube events and the Fermi gamma
ray excess can be simultaneously explained in the \textit{MSSM-like} model
only when the pseudoscalar $a$ is heavy ($m_a \gtrsim m_h/2$).

{\bf Flipped model.} The second model, which we call \textit{Flipped} is a
flipped Two Higgs Doublet Model~\cite{Branco:2011iw, PhysRevD.41.3421,
Grossman:1994jb, Akeroyd:1996he, Aoki:2009ha, PhysRevD.30.1529}. This means
that one Higgs doublet couples to up quarks and leptons, while the other
couples to down quarks.  The difference between this model and the
\textit{MSSM-like} model is that the coupling to leptons in the
\textit{Flipped} scenario is proportional to $\cot{\beta}$ rather than
$\tan\beta$ and is thus suppressed rather than enhanced in the large
$\tan{\beta}$ region. Therefore, limits from the lepton sector will be
significantly weaker. The couplings to up-type quarks and down-type quarks are
the same as in the \textit{MSSM-like} model.

{\bf Vector-like quark model.} The third model has no extended Higgs sector,
and the pseudoscalar mediator $a$ does not directly couple to SM quarks. Instead,
it couples to new, heavy vector-like quarks, which in turn mix with the SM
quarks~\cite{Izaguirre:2014vva}. Since $a$ has no couplings to leptons in this
model and since there is no extended Higgs sector, we expect constraints to be
weaker than in the other two scenarios. However, the mass of the heavy
vector-like quark should be large to avoid LHC limits.

\section{Boosted dark matter in IceCube}
\label{sec:ic}

\subsection{Primary signal: scattering of boosted DM particles on nuclei}
\label{sec:icecube-DM}

\begin{figure*}
  \begin{tabular}{c@{\qquad\qquad}c}
    \includegraphics[height=3.5cm]{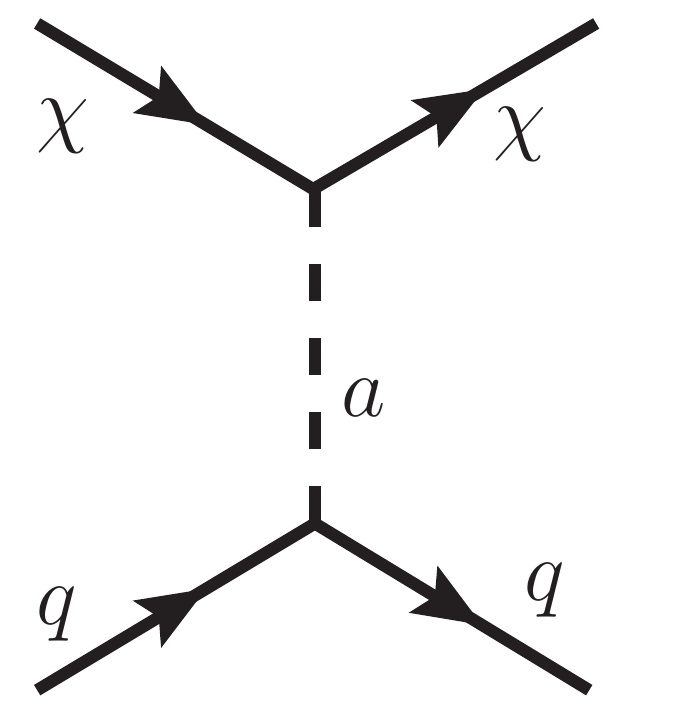} &
    \includegraphics[height=3.5cm]{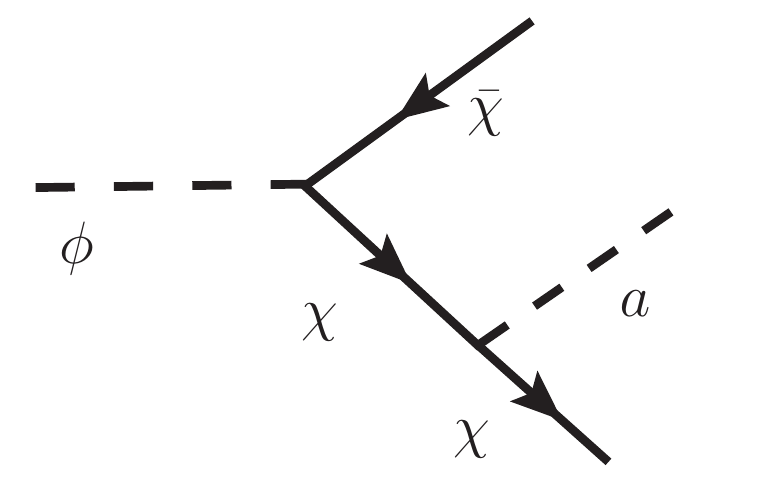} \\
    (a) & (b)
  \end{tabular}
  \caption{The Feynman diagrams for (a) the scattering of light DM particle $\chi$
    on nucleons and (b) the 3-body decay $\phi \to \chi \bar{\chi} a$, which
    produces a flux of high energy pseudoscalars whose decay products contribute to
    astrophysical neutrino, gamma ray and positron fluxes.}
  \label{fig:icecube-feyn}
\end{figure*}

Highly boosted $\chi$ particles from the DM decay process $\phi \to \chi
\bar\chi$ can scatter on atomic nuclei in the IceCube detector through their
coupling to the pseudoscalar mediator $a$ (see fig.~\ref{fig:icecube-feyn} (a)).
At the high energies we are interested in, the scattering is deep inelastic.
Phenomenologically, this process is very similar to neutral
current scattering of neutrinos, hence its characteristic signature is a
shower-like event topology. The deposited (or visible) energy $E_\text{dep}$ in
this case is the energy of the recoil nucleus or its fragments.

The total number of shower events from $\chi$ scattering in a given
$E_\text{dep}$ bin $[E_\text{dep}^\text{min}, E_\text{dep}^\text{max}]$ is
given by \cite{Mena:2014sja}
\begin{align}
  N_\chi^{\rm sh,NC} = T \int_{E_\chi^\text{min}}^{m_{\phi}/2} \! dE_\chi \,
                   \frac{d\Phi_\chi}{d E_\chi} \times
                   \int_{E_\text{dep}^\text{min}}^{E_\text{dep}^\text{max}} \! dE_\text{dep}
                   \frac{M^\text{NC}(E_\text{dep})}{18 m_N}
                   \bigg(
                     10 \frac{d\sigma_p(E_\chi,E_\text{dep})}{dE_\text{dep}} +
                      8 \frac{d\sigma_n(E_\chi,E_\text{dep})}{dE_\text{dep}}
                   \bigg) \,.
  \label{eq:Nchi}
\end{align}
Here, $T$ is the observation time, $m_N$ is the nucleon mass,
$d\sigma_{p (n)} / dE_\text{dep}$ is the differential
scattering cross section on protons (neutrons). $M^\text{NC}(E_\text{dep})$ is the
effective detector mass of IceCube for neutral current scattering as a function
of $E_\text{dep}$. Details on how we estimate $M^\text{NC}(E_\text{dep})$ from the
effective detector mass as a function of incoming neutrino energy, $M^\text{NC}(E_\nu)$,
published by the IceCube collaboration~\cite{Aartsen:2013jdh} are given in
appendix~\ref{app:meff}. Our estimate of $M^\text{NC}(E_\text{dep})$ is in agreement
with the results from ref.~\cite{Palomares-Ruiz:2015mka}, which found
that $M^\text{eff}(E_\text{dep})$ is universal for NC and CC interactions.

The flux of boosted light DM particles $\chi$
has a galactic component $\Phi_\chi^\text{GC}$
and an extragalactic component $\Phi_\chi^\text{EG}$:
\begin{align}
  \frac{d\Phi_\chi}{d E_\chi} = \frac{d\Phi_\chi^\text{GC}}{d E_\chi}
                              + \frac{d\Phi_\chi^\text{EG}}{d E_\chi} \,.
\end{align}
The galactic contribution
is given by~\cite{Esmaili:2012us},
\begin{align}
  \frac{d\Phi_\chi^\text{GC}}{dE_\chi}
    &= \int\!d\Omega_\psi \,
       \frac{1}{4\pi m_\phi \tau_\phi} \frac{dN_\chi}{dE_\chi}
       \int_{los} \! ds \, \rho_\text{halo}
       \big( \vec{r}(s, \psi) \big) \,,                     \label{eq:DMflux0} \\[0.2cm]
    &= 2.1 \times 10^{-10}\ \text{cm$^{-2}$ sec$^{-1}$} \, \times
       \bigg( \frac{10^{26}~ \text{sec}}{\tau_\phi} \bigg)
       \bigg( \frac{1~ \text{PeV}}{m_\phi} \bigg)
       \bigg( \frac{dN_\chi}{dE{\chi}} \bigg) \,,                \nonumber
\end{align}
Here, $m_\phi$ and $\tau_\phi$ are the mass and lifetime of
the heavy DM particle $\phi$, respectively,
$\rho_\text{halo}$ is the DM density distribution in the
Milky Way, $\vec{r}(s,\psi)$ is the position vector relative to the origin at the
galactic center, $s$ is the distance along the line of sight and $\psi$ is its
angular direction. We integrate the flux over the solid angle
$\Omega_\psi$ and integrate along the line of the sight $s$.
The energy spectrum of boosted $\chi$ particles is simply
$dN_\chi / dE_\chi = \delta(E_\chi - m_\phi/2)$. The spectrum of
antiparticles, $dN_{\bar{\chi}} / dE_{\bar{\chi}}$ is the same.
The extragalactic contribution to the flux of $\chi$ particles is~\cite{Esmaili:2012us}
\begin{align}
  \frac{d\Phi_\chi^\text{EG}}{dE_\chi}
    &= \frac{\Omega_\text{DM} \rho_c}{m_\phi \tau_\phi}
       \int_0^\infty \! dz \, \frac{1}{H(z)} \, \frac{dN_\chi}{dE_\chi}[(1+z) E_\chi]
                     \,.        \label{eq:DMflux2}
\end{align}
In this expression, $H(z) \simeq H_0 \sqrt{\Omega_\Lambda + \Omega_\text{m} (1
+ z)^3}$ is the Hubble expansion rate as a function of redshift $z$. It depends on
the Hubble constant $H_0 = H(0)$, the dark energy density $\Omega_\Lambda
\sim 0.692$ and the matter density $\Omega_\text{m} \sim 0.308$.
The cold dark matter density
$\Omega_\text{DM}$ is $0.258$, and the critical density
of the Universe $\rho_c$ is given by $\rho_c \simeq 4.9 \times 10^{-6}\
\text{GeV}/\text{cm}^3$~\cite{Planck:2015xua}.  Note
that we do not account here for attenuation of the $\chi$ flux due to
scattering on the interstellar and intergalactic medium.  This attenuation is
already small for neutrinos~\cite{Esmaili:2012us}, and the $\chi$ scattering
cross section is even smaller than the neutrino scattering cross section.

The differential cross section for $\chi$ scattering on a proton $p$ (neutron $n$)
of mass $m_N$ is
\begin{align}
  \frac{d\sigma_{p (n)}}{dx \, dE_\text{dep}}
    &= \sum_q \frac{1}{32\pi s} \,
       \frac{4 s x m_N}{(s - m_\chi ^2 - x^2 m_N^2)^2 - 4 x^2 m_N^2 m_\chi^2} \times
       f_q^{p (n)}(x) \times
       \frac{1}{4} \sum_\text{spins} |\mathcal{M}_q|^2 \label{eq:scatteringeqs} \\
\intertext{with}
  \frac{1}{4} \sum_\text{spins} |\mathcal{M}_q|^2
    &= \frac{2 g_\chi^2 g_{Y_q}^2 m_q^2 (Q^2)^2}{v^2 (Q^2 + m_a^2)^2} \,, \label{eq:scatteringeqs2}
\end{align}
where $x$ is the Bjorken scale variable, $s = m_\chi^2 + x^2 m_N^2 + 2 x m_N E_\chi$
is the center of mass
energy, and $Q^2 = 2x m_N E_{\rm dep}$ is the momentum transfer in the
scattering.  $E_{\chi}$ is the energy of the incoming particle $\chi$ and the
nucleon is assumed to be at rest initially. $E_{\rm dep}$ is the energy
transferred to the hadronic system in the lab frame during the scattering. We
are interested in events with a large deposited energy $E_{\rm dep} \gtrsim
10$~TeV in this analysis due to the IceCube energy threshold.

The factor $f_q^{p (n)}(x)$ in eq.~\eqref{eq:scatteringeqs} is the parton distribution
function (PDF) for protons (neutrons) and quark flavor $q$.  We use the PDFs from
NNPDF3.0~\cite{Ball:2008by}, which are valid in the range $x \in [10^{-9}, 1]$
and $Q^2 \in [2~\text{GeV}^2, \,10^8~\text{GeV}^2]$ and contain the most recent
deep inelastic scattering data.\footnote{At $x$ very close to 1, the NNPDF3.0
PDFs are not smooth. Even though the
large $x$ region is not important for our results, we do not use NNPDF at $x >
0.1$, but use CTEQ5~\cite{Lai:1999wy} PDFs instead.} In the calculation, we set
the PDFs equal to $0$ when $Q^2$ is smaller than $2~\text{GeV}^2$. Because the
cross section is proportional to $1/(Q^2+m_a^2)^2$, it becomes large when $Q^2$
is small. Our cutoff at low $Q^2$ would therefore affect the results for $m_a
\lesssim$~GeV.  In the following, however, we focus on the mass range $m_a >
10$~GeV, and we have checked that in this case the contribution of the $Q^2 <
2~\text{GeV}^2$ region to the cross section is negligible. If one is interested
in an extremely light $t$-channel mediator with $m_a^2 \ll 2\text{GeV}^2$, then
the $Q^2 \ll 2\text{GeV}^2$ and $x \ll 1$ region, corresponding to exchange of
a nearly on-shell $a$, is important. In this region, the PDF description breaks
down and one should instead calculate the cross section for $a^{*}$ absorption
by protons along the lines of the equivalent photon approximation in
deep-inelastic scattering of electrons on protons.  We have used the central
values of the PDFs, but have checked that varying them within the error band
changes the total cross section by only $\mathcal{O}(10\%)$. The impact on the
energy dependence of the differential cross section is also negligible.

\subsection{Secondary signal: neutrino flux from 3-body decays of heavy DM}
\label{sec:icecube-neutrino}

\begin{figure*}
  \includegraphics[width=0.45\textwidth]{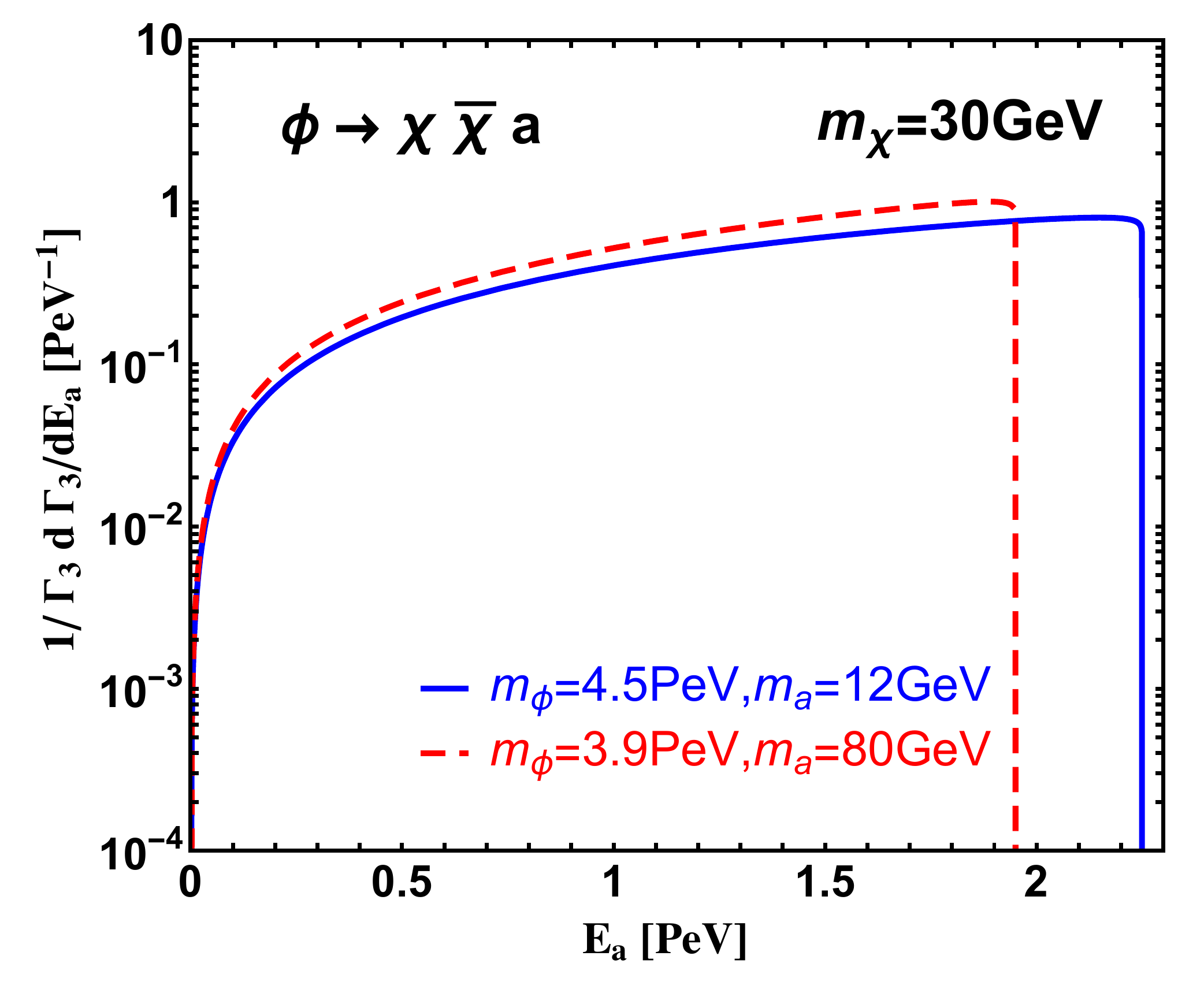}
  \caption{The energy distribution of pseudoscalar particles $a$ produced
    in the 3-body decay $\phi \to \chi \bar\chi a$. The parameter values
    have been fixed at our benchmark values $m_\phi = 4.5~(3.9)$~PeV, $m_\chi = 30$~GeV,
    $m_a = 12\ \text{GeV} (80\ \text{GeV})$ for the solid blue (dashed red)
    lines.}
    \label{fig:Eadistribution}
\end{figure*}

As mentioned in the Introduction, the boosted DM scenario predicts not only a
population of high energy events from the scattering of boosted $\chi$
particles, but also a contribution at lower energy from neutrinos produced in
the 3-body decay $\phi \to \chi\bar\chi a$ (see fig.~\ref{fig:icecube-feyn} (b)),
followed for instance by $a \to b \bar{b}$.  Since boosted DM can only
explain the IceCube events if $\chi$ particles can scatter on nucleons,
a mediator particle like $a$ is always needed and the existence of the
3-body decay process is thus very generic. Making $a$ heavy does not
significantly influence the 3-body decay rate unless $m_a$ becomes comparable
to $m_\phi$. The differential decay width of the 3-body decay is,
in the limit $m_a \to 0$ and at leading order in $m_\chi$,
\begin{align}
  \frac{d\Gamma_3(\phi \to \chi \bar{\chi} a)}{dE_a}
    &= \frac{g_\chi^2 y_{\phi\chi}^2 E_a}{16 \pi^3 m_\phi}
       \log\bigg( \frac{m_\phi^2 - 2 m_\phi E_a}{m_\chi^2} \bigg) \,,
  \label{eq:Gamma-3body}
\end{align}
where $E_a$ is the energy of $a$ in the rest frame of $\phi$.
The branching ratio is
\begin{align}
  \BR_3(\phi \to \chi \bar{\chi} a) &=
    \frac{\Gamma_3(\phi \to \chi \bar{\chi} a)}
         {\Gamma_3(\phi \to \chi \bar{\chi} a) + \Gamma_2(\phi \to \chi \bar{\chi})} \\
  &\simeq \frac{\Gamma_3(\phi \to \chi \bar{\chi} a)}
               {\Gamma_2(\phi \to \chi \bar{\chi})} \,,
  \label{eq:br3}
\end{align}
where
\begin{equation}
  \Gamma_2(\phi \to \chi \bar\chi) = \frac{y_{\phi\chi}^2 m_\phi}{8\pi}
    \bigg(1 - \frac{4 m_\chi ^2}{m_\phi^2} \bigg)^{3/2}
\end{equation}
is the rate of the dominant 2-body decay.  In the second line of
eq.~\eqref{eq:br3}, we have assumed that $g_\chi$ is small so that the 3-body
decay width is much smaller than the 2-body decay width. We can see from
Table~\ref{tab:benchmarks} that this assumption is satisfied at our benchmark
points. We plot the energy spectrum of $a$ particles from 3-body decay of
$\phi$ in fig.~\ref{fig:Eadistribution}.

The decay of $a$ to light quarks or $b$ quarks produces neutrinos after
parton showering, hadronization and hadron decay.  We take the
spectra of the secondary neutrinos from each $a$ decay in the $a$ rest frame
from \cite{Cirelli:2010xx} and boost them
into the laboratory frame by folding with the $E_a$ distribution
from fig.~\ref{fig:Eadistribution}~\cite{Liu:2014cma}.
Multiplying by $\BR_3(\phi \to \chi \bar\chi a)$ gives us the
number $dN_\nu / dE_\nu$ of neutrinos per energy interval $dE_\nu$ per
$\phi$ decay.
The flux of secondary neutrinos is then obtained from equations very similar to
eqs.~\eqref{eq:DMflux0} and \eqref{eq:DMflux2} by simply replacing the factor
$dN_\chi / dE_\chi$ by $dN_\nu / dE_\nu$.  The strength of the indirect signal
is proportional to $g_\chi^2 f_\phi / \tau_\phi$ once the masses $m_\phi$, $m_\chi$
and $m_a$ are fixed.  In principle, one might also include a factor of the form
$\exp[-\text{Abs}(E_\chi, z)]$ in the expression for the extragalactic flux to
account for the absorption of neutrinos in interactions with the
cosmological relic neutrino background and with the intergalactic
medium~\cite{Esmaili:2012us}.
However, these effects are negligible in our analysis and we therefore do not
include such an attenuation factor.  Moreover, the high energy neutrino flux
reaching the detector from below is affected by neutrino interactions during
passage through the Earth.  In particular, at energies above $\sim 100$~TeV,
the neutrino--nucleon interaction cross section is so large that the Earth can
attenuate the neutrino flux.  On the other hand, electron and muon neutrinos
can be regenerated in the decay of tau leptons produced in $\nu_\tau$ CC
interactions.  The net effect of both absorption and regeneration is a reduction of
the neutrino flux by about 15\% at neutrino energies $\sim
100$~TeV~\cite{Palomares-Ruiz:2015mka}, and we therefore neglect this small
effect in our calculation.

\begin{figure*}
  \includegraphics[width=0.45\textwidth]
    {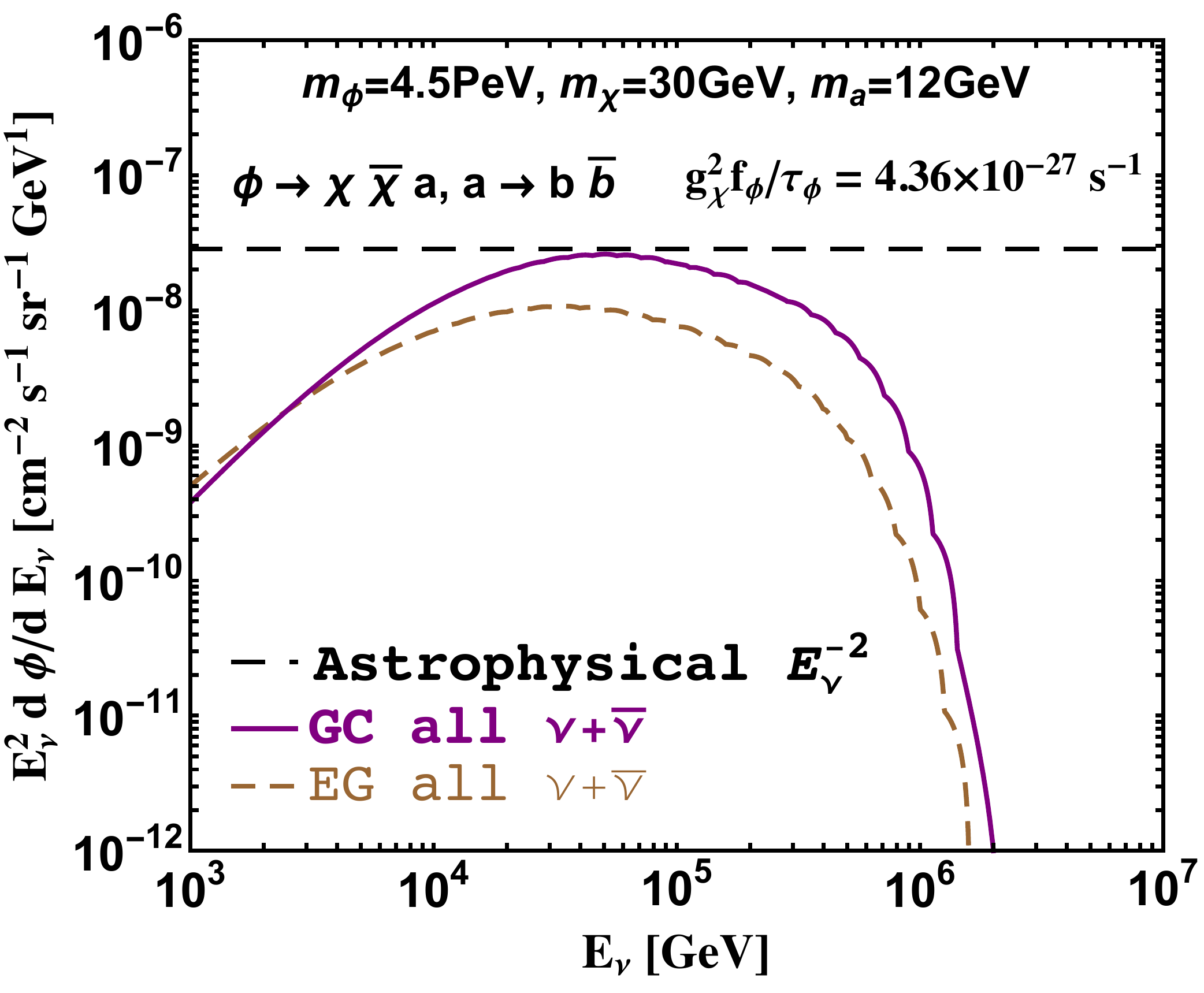} %
  \includegraphics[width=0.45\textwidth]
    {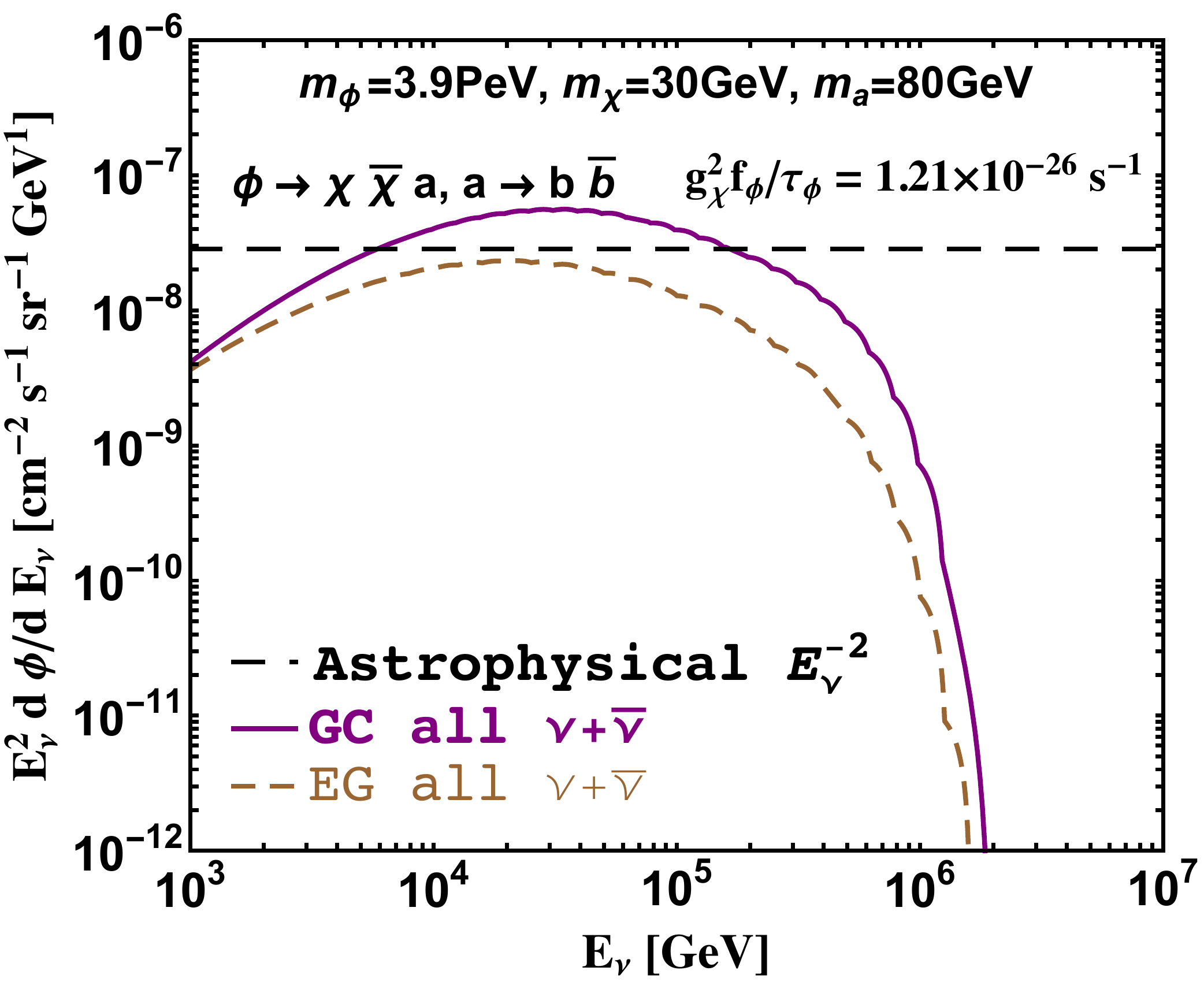} %
  \caption{The galactic and extragalactic neutrino fluxes from
    3-body decay of heavy DM, $\phi \to \chi \bar\chi a$, followed
    by $a \to b \bar{b}$. We have added up the neutrino and antineutrino
    fluxes and have also summed over neutrino flavors. The horizontal dashed line
    shows the generic flux expected from astrophysical sources,
    $E_\nu^{-2}$, normalized such that optimum agreement with the IceCube data
    is achieved~\cite{Aartsen:2014gkd}.
    The model parameters are set to the benchmark values given in the plot.}
  \label{fig:neutrinoflux}
\end{figure*}

We plot the expected contributions to the neutrino flux from galactic and
extragalactic $\phi \to \chi \bar\chi + (a \to b \bar{b})$ decays in
fig.~\ref{fig:neutrinoflux} for our two benchmark points.  Since the
neutrinos originate mostly from meson decays after hadronization of the $b$
quarks, their flavor ratio after propagation is naturally $(1:1:1)_E$.
Therefore, we have summed the different flavors, as well as the neutrino and
antineutrino fluxes, in fig.~\ref{fig:neutrinoflux}.  We see that the
secondary neutrinos are softer by about one order of magnitude compared to the
boosted DM particles $\chi$.  The extragalactic flux is in general softer than
the galactic one due to redshift.

Note that, besides the secondary neutrino flux, there is also a population of
boosted DM events from scattering of the $\chi$ particles produced in 3-body
decay $\phi \to \chi \bar{\chi} a$.  We neglect these events for
the following reasons: first, the 3-body branching ratio is almost two orders of
magnitude smaller than the 2-body branching ratio. Second, the spectrum of
$\chi$ particles from 3-body decays is softer than the one from 2-body decays
and would therefore contribute only in a regime with larger expected
backgrounds. Third, a 3-body decay $\phi \to \chi \bar{\chi} a$ produces only
two $\chi$ particles, but typically more than two neutrinos~\cite{Cirelli:2010xx}.
Thus, the flux of $\chi$ particles from 3-body decay is subdominant
compared to the secondary neutrino flux. Fourth, the $\chi$ scattering cross
section on nucleons is usually smaller than the neutrino charged current cross
section.

\subsection{Fitting procedure}
\label{sec:icecubce-fit}

To determine the preferred parameter regions for the boosted DM scenario,
we use the log likelihood ratio (LLR) method. The LLR is defined as follows:
\begin{align}
  &\text{LLR} \bigg(m_\phi, \frac{g_{Y_b}^2 g_\chi^2 f_\phi}{\tau_\phi},
                           \frac{g_\chi^2 f_\phi}{\tau_\phi} \bigg)  \nonumber\\
    &\qquad = \log \left(
        \frac{ \mathop\text{Max}\limits_{x \in [-\infty,\infty]}
               \Big[
                  f_\text{Gauss}(x) \,
                  \prod_i f_\text{Poisson} \Big(
                  S_i \Big(m_\phi, \frac{g_{Y_b}^2 g_\chi^2 f_\phi}{\tau_\phi},
                           \frac{g_\chi^2 f_\phi}{\tau_\phi} \Big)
                + B_i + x \,\Delta B_i \, \Big| \, O_i \Big) \Big] }
             { \mathop\text{Max}\limits_{x' \in [-\infty,\infty]}
               \Big[
                 f_\text{Gauss}(x') \,
                 \prod_i f_\text{Poisson} \big(
                 B_i + x' \,\Delta B_i \big| O_i \big) \Big] }
      \right) \,.
  \label{eq:chi2-fit}
\end{align}
Here, $f_\text{Poisson}(\mu|n) = \mu^n e^{-\mu} / n!$ is the Poisson
likelihood function and $S_i(m_\phi, g_{Y_b}^2 g_\chi^2 f_\phi/\tau_\phi,
g_\chi^2 f_\phi/\tau_\phi)$,
$B_i$ and $O_i$ are the predicted signal event rate, the predicted background event
rate, and the observed event rate in the $i$-th energy bin, respectively.
$\Delta B_i$ is the $1\sigma$ error on the background prediction.  When the
nuisance parameter $x$ is $1$ ($-1$), the error $x \, \Delta B_i(x)$ describes the
upper (lower) limits of the error band, and when $x=0$ the background takes its
central value.  The term $f_\text{Gauss}(x)$ corresponds to a normal distribution
in $x$ and is the Gaussian pull term for the nuisance parameter $x$.
By using only one nuisance parameter, we effectively assume
that the background uncertainty is correlated between bins.

\subsection{Results}
\label{sec:icecube-results}

\begin{figure*}
  \begin{tabular}{ccc}
    \includegraphics[width=0.48\textwidth]
      {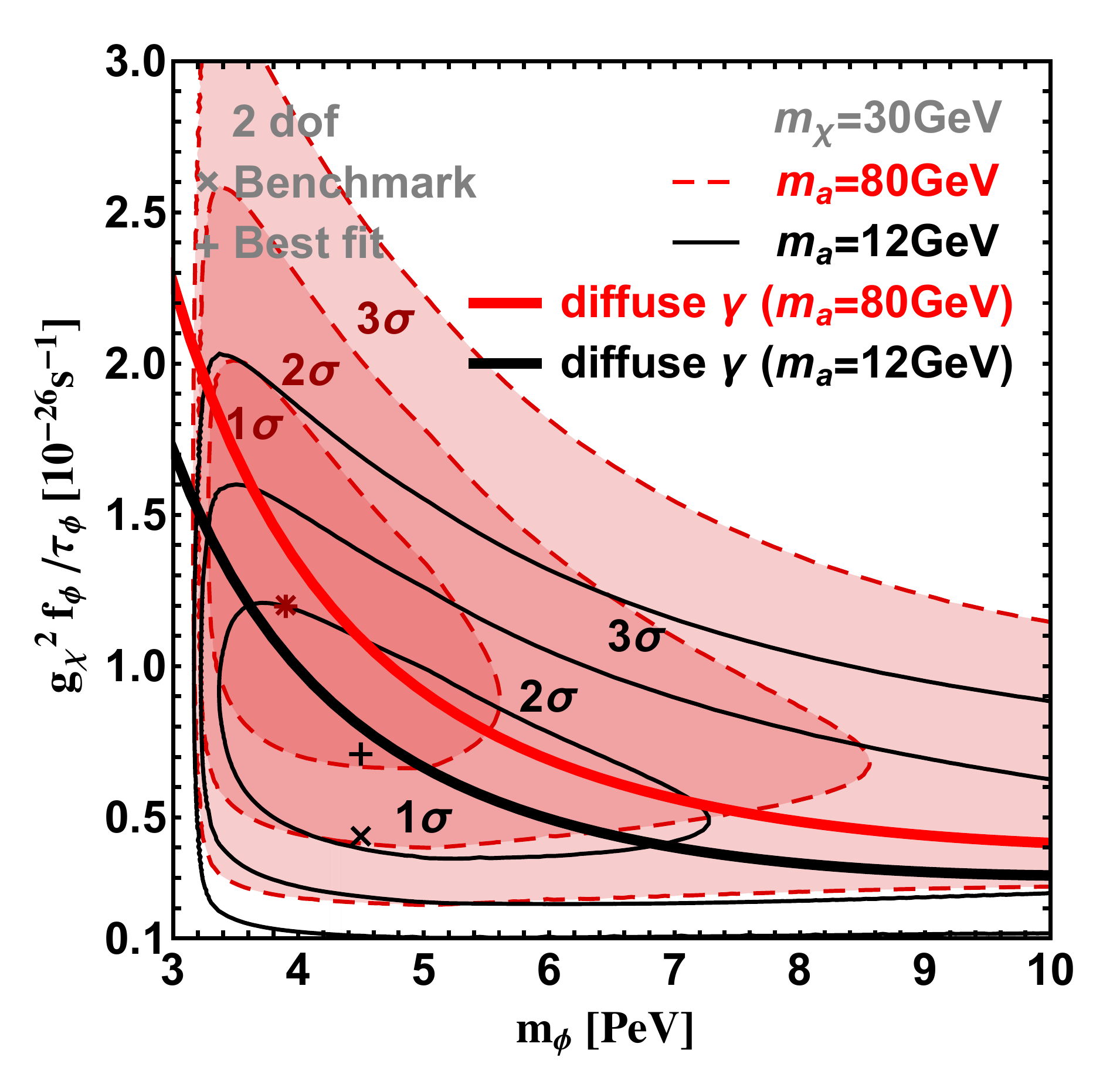} &
    \includegraphics[width=0.48\textwidth]
      {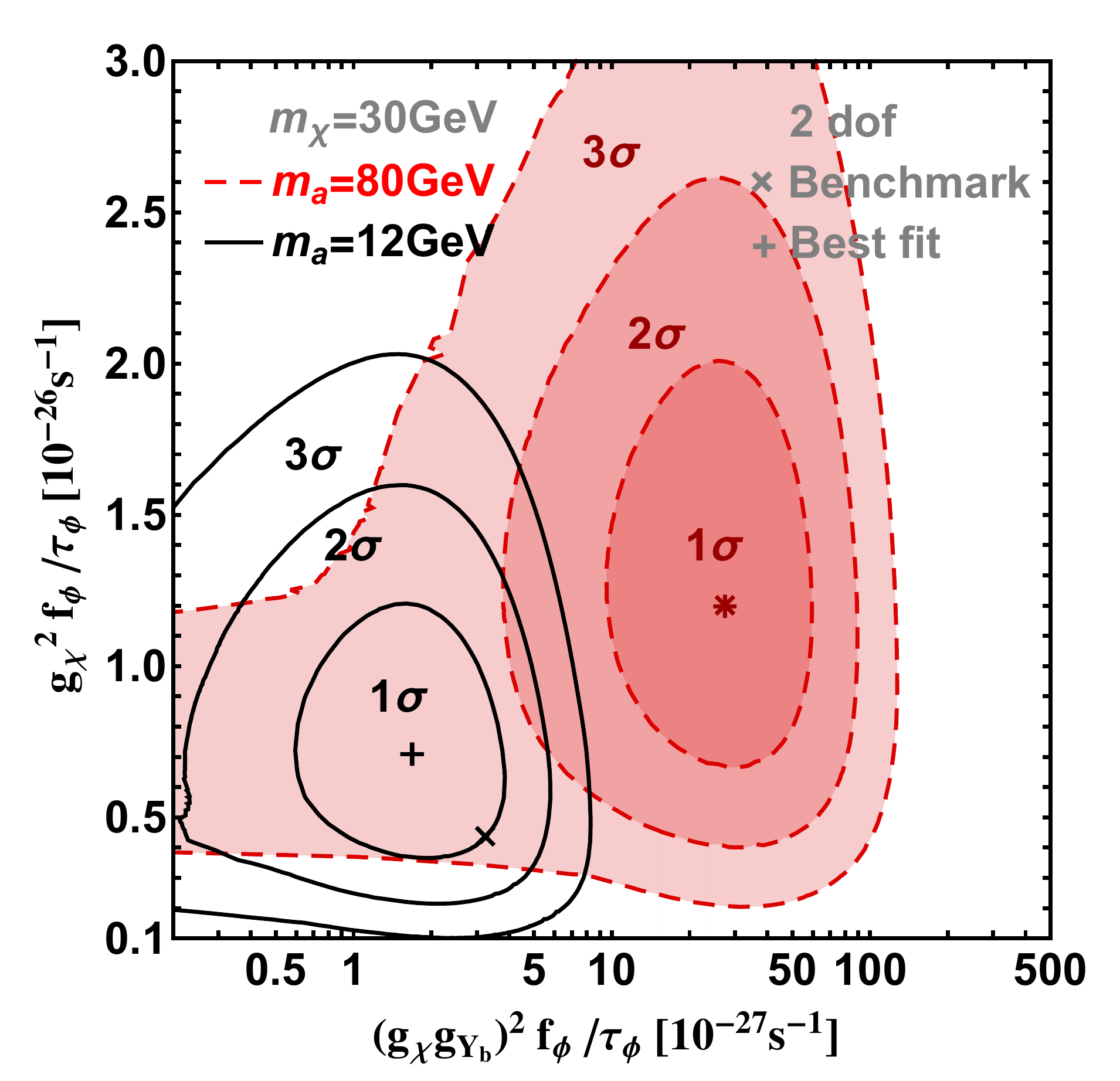} \\
    \includegraphics[width=0.48\textwidth]
      {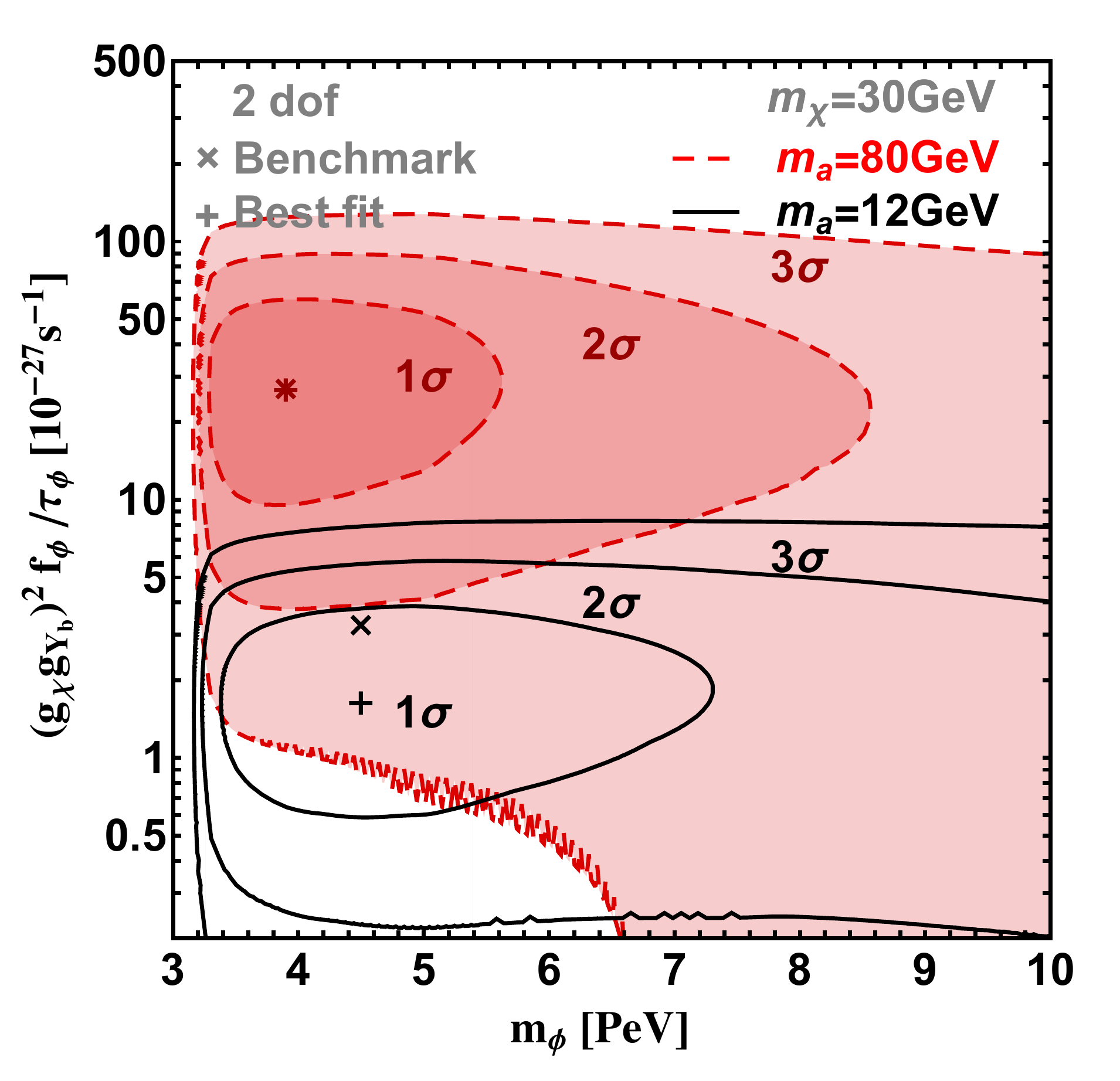} &
  \end{tabular}
  \caption{Preferred parameter regions for the boosted DM scenario from our fit to
    IceCube high energy data~\cite{Aartsen:2014gkd}.  The three panels show
    2-dimensional projections of the 3-dimensional parameter space spanned by
    the heavy DM mass $m_\phi$, the product $g_\chi^2 g_{Y_b}^2 f_\phi /
    \tau_\phi$ to which the scattering rate of boosted $\chi$ particles is
    proportional, and the combination $g_\chi^2 f_\phi / \tau_\phi$ to which
    the flux of secondary neutrinos is proportional. (Here, $g_\chi$ and
    $g_{Y_b}$ are coupling constants, $f_\phi$ is the cosmological abundance of
    $\phi$, and $\tau_\phi$ is its lifetime.) Solid black unshaded (red dashed
    shaded) contours show the preferred parameter regions at $1,~2,~3\sigma$
    for $m_a = 12$~GeV ($m_a = 80$~GeV) and the black (red) ``$+$'' signs
    indicate the best fit points.  At $m_a = 80$~GeV, the best fit point is
    identical to one of our benchmark points (red ``$\times$'' sign) from
    table~\ref{tab:benchmarks}, while for $m_a = 12$~GeV we define our
    benchmark point (black ``$\times$'' sign) slightly away from the best fit.
    This way, both benchmark points can also explain the galactic center gamma
    ray excess and evade all constraints.  In the upper left hand plot we also
    show as a thick black (thick red) curve the strongest exclusion limits on the
    $m_a = 12$~GeV ($m_a = 80$~GeV) benchmark model, coming from diffuse
    $\gamma$ ray searches (see sec.~\ref{sec:gamma}).  We use $m_\chi = 30$~GeV
    for the mass of the light, boosted, DM particle here, motivated by the
    galactic center gamma ray excess, but note that $m_\chi$ does not affect the
    IceCube event rate as long as $m_\chi \ll m_\phi$.}
  \label{fig:contours}
\end{figure*}

\begin{figure*}
  \begin{tabular}{c@{\quad}c}
    \includegraphics[width=0.48\textwidth]{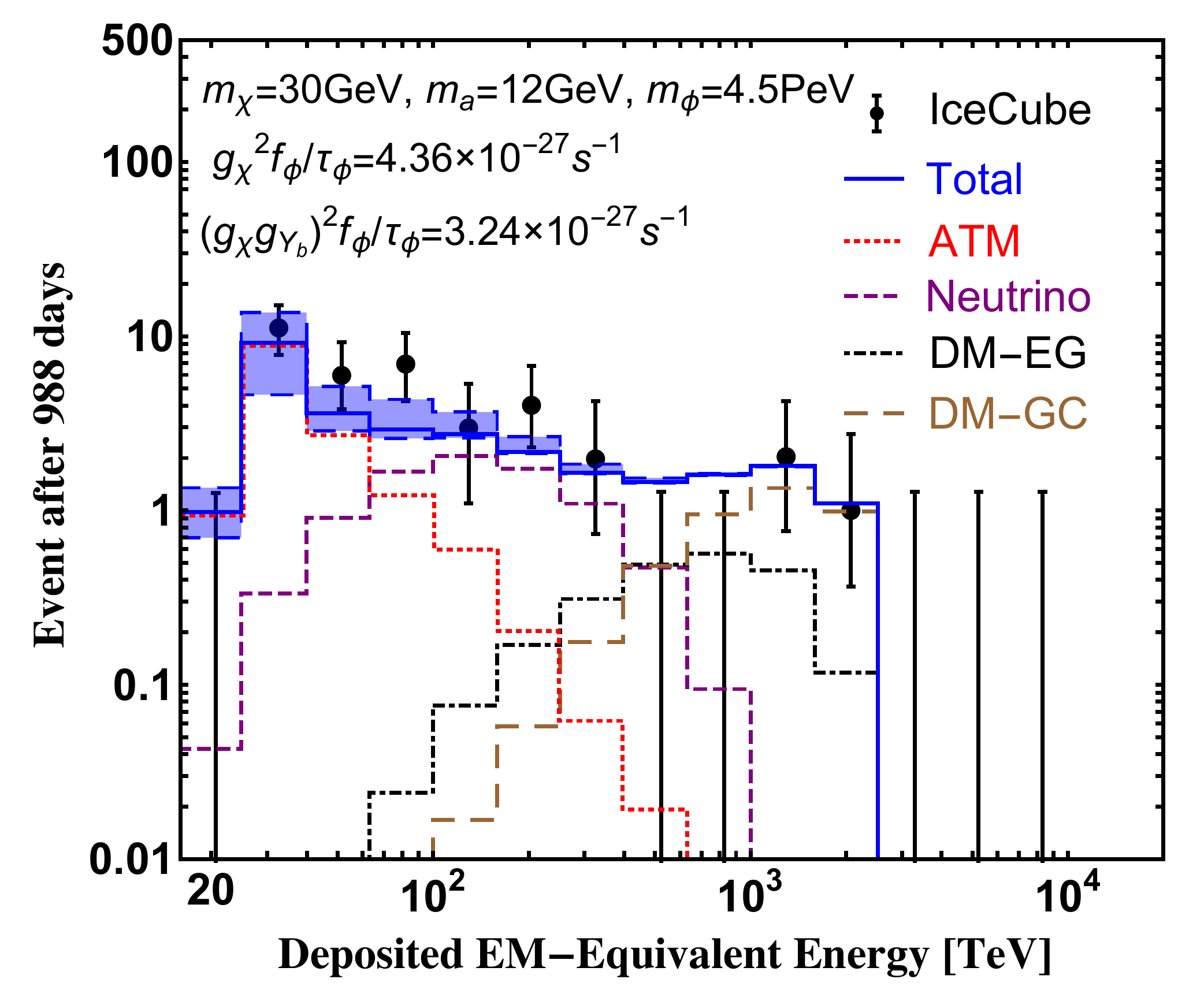} &
    \includegraphics[width=0.48\textwidth]{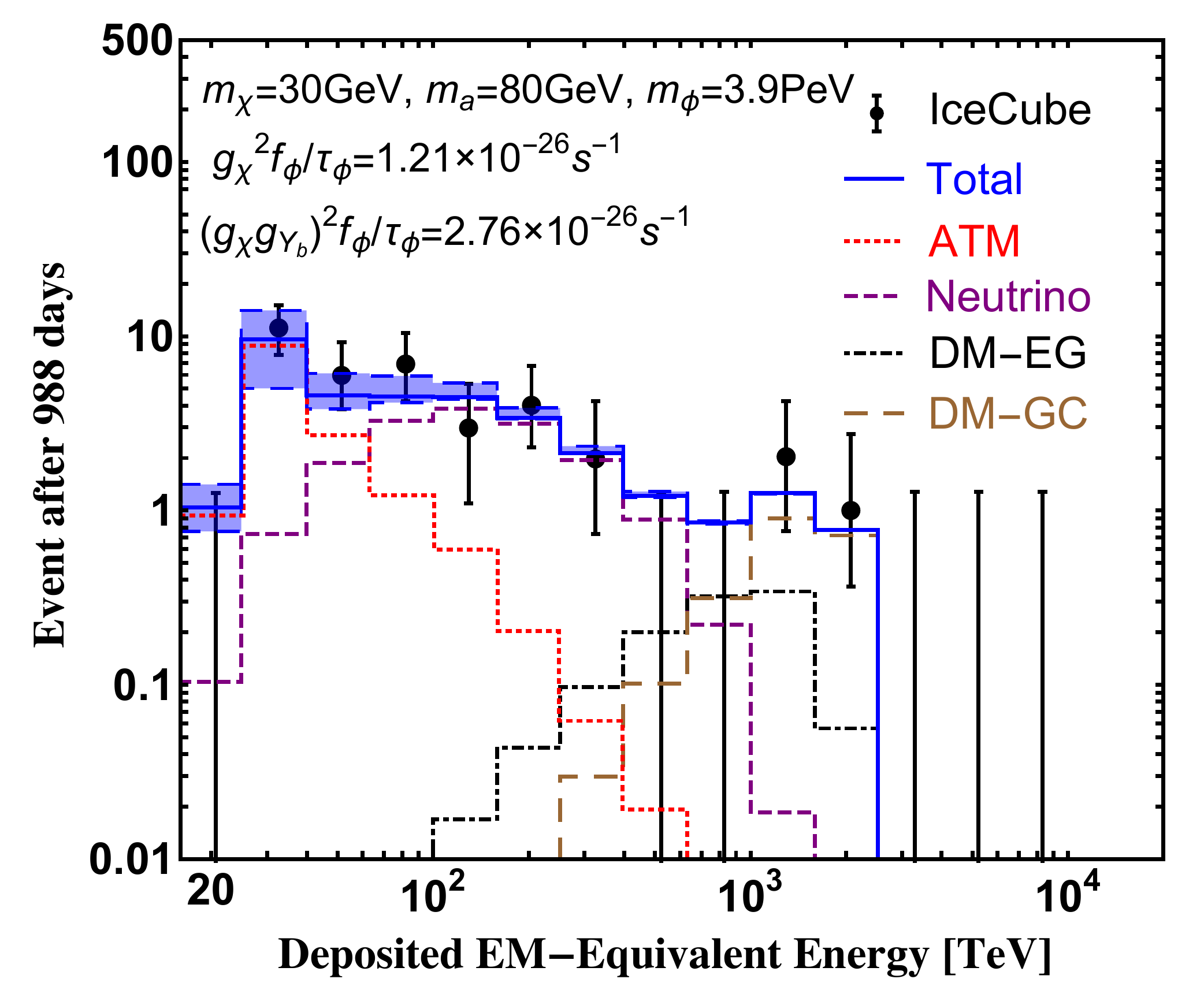} \\
    (a) & (b)
  \end{tabular}
  \caption{Comparison of IceCube high energy data~\cite{Aartsen:2014gkd}
    to the prediction at our two benchmark points (see
    Table~\ref{tab:benchmarks}.  We plot the signals from galactic (brown
    dashed) and extragalactic (black dot-dashed) $\phi \to \chi \bar\chi$
    decays, as well as the contribution from secondary neutrinos produced in
    $\phi \to \chi \bar\chi + (a \to b \bar{b})$ (purple dashed) separately.
    The red dotted lines show the atmospheric neutrino background (``ATM''),
    the blue bars depict the background uncertainty and the solid blue lines
    show the total expected event rate.  We have taken the mass of the
    pseudoscalar mediator $m_a$ to be 12~GeV (80~GeV) in the \textit{left panel}
    (\textit{right panel}).  We always use $m_\chi = 30$~GeV for the mass of
    the light (boosted) DM particle here, motivated by the galactic center gamma
    ray excess, but note that $m_\chi$ does not affect the IceCube event rate
    as long as $m_\chi \ll m_\phi$.}
  \label{fig:IceCubeSpectrum}
\end{figure*}

We show the results of our fit in fig.~\ref{fig:contours} and compare the best
fit points to the IceCube data in fig.~\ref{fig:IceCubeSpectrum}.  For the
mediator mass $m_a = 12$~GeV (80~GeV), the three panels of
fig.~\ref{fig:contours} give the best fit points (black (red) ``$+$'' signs)
and preferred parameter regions (black unshaded contours (red shaded contours))
at $1,~2,~3\sigma$ confidence level.  For $m_a = 80$~GeV, the best fit point,
marked by a red ``$\times$'' sign, corresponds to one of our benchmark points
from table~\ref{tab:benchmarks}, while for $m_a = 12$~GeV, the benchmark point
(indicated by the black ``$\times$'' sign) is slightly shifted compared to the
best fit in order to be consistent also with the galactic center excess and
with all constraints. The larger value of $m_a$ is particularly interesting for
the \textit{MSSM-like} and \textit{Flipped} models, where it helps to evade
important constraints from $B_s \to \mu^+\mu^-$ decays and from $h \to a a$
decays. (see sec.~\ref{sec:flavor-collider}).  Note that we parameterize the
parameter space in fig.~\ref{fig:contours} in terms of three parameters: the
heavy DM mass $m_\phi$; the combination $g_{Y_b}^2 g_\chi^2 f_\phi / \tau_\phi$
of the $a$ coupling constants, the cosmological abundance $f_\phi$ of the heavy
DM particle $\phi$ and its lifetime $\tau_\phi$, to which the $\chi$ scattering
rate is proportional; and the ratio $g_\chi^2 f_\phi / \tau_\phi$ to which the
interaction rate of secondary neutrinos is proportional.  In the upper left
hand plot, we also show constraints from the diffuse $\gamma$ ray flux (see
sec.~\ref{sec:gamma}) as thick black (red) lines. We always fix the mass of the
light DM particle at $m_\chi = 30$~GeV, as motivated by the galactic center
gamma ray excess, see sec.~\ref{sec:GCE}.  As expected, the best fit point is
always around $m_\phi \sim 4$~PeV due to the lack of IceCube events above
2~PeV.  In fig.~\ref{fig:IceCubeSpectrum}, we compare the IceCube data from
ref.~\cite{Aartsen:2014gkd} to our predictions at the benchmark points.  We
also show the individual contributions to the spectrum separately: the
atmospheric (``ATM'') neutrino background (red dotted), the galactic (brown
dashed) and extragalactic (black dot-dashed) fluxes of boosted $\chi$
particles, and the flux of secondary neutrinos from $\phi \to \chi\bar\chi + (a
\to b \bar{b})$ decay (purple dashed).

We see that both the galactic and extragalactic $\chi$ fluxes contribute at PeV
energies, with the latter being somewhat softer due to redshift.  Actually, the
integrated fluxes of the two components are comparable, but since the
scattering cross-section is higher when the energy of the incoming $\chi$
particle is larger, the softer component is subleading experimentally.  Below
1~PeV, the boosted DM event rates drop because of the $Q^2$ dependence of the
scattering matrix element, eq.~\eqref{eq:scatteringeqs2}.  In their place, the
secondary neutrino flux takes over below $\sim 500$~TeV, so that a good fit to
the IceCube data is obtained at all energies.  Note that the normalization of
the secondary neutrino flux is set by the parameter combination
$g_\chi^2 f_\phi / \tau_\phi$ and is thus not directly correlated with the boosted DM
scattering rate, which is proportional to $g_{Y_b}^2 g_\chi^2 f_\phi / \tau_\phi$.

Comparing our two benchmark values of $m_a$ (shaded vs.\ unshaded contours in
fig.~\ref{fig:contours}, left vs.\ right panel in
fig.~\ref{fig:IceCubeSpectrum}), we observe that the choice of $m_a$ has a
small influence on the spectral shape of the DM contributions, but its main
impact is on the overall rate.  Therefore, at larger $m_a$, the best fit value
of $g_{Y_b}^2 g_\chi^2 f_\phi / \tau_\phi$ is significantly larger than at
smaller $m_a$.  When $a$ is heavy, one either needs large $g_{Y_b} g_\chi$
coupling to keep the scattering cross section of the boosted DM particle $\chi$
on nucleons unchanged, or the flux of $\chi$ particles must be enhanced by
decreasing the heavy DM lifetime $\tau_\phi$.
Note that the two benchmark models shown in
figs.\ref{fig:contours} and \ref{fig:IceCubeSpectrum} explain not only the
IceCube data, but also the galactic center gamma ray excess (see
sec.~\ref{sec:GCE}) and are consistent with all constraints (see
sec.~\ref{sec:constraints}).

An interesting aspect of our boosted DM scenario is that a dip in the event
spectrum is predicted between recoil energies of $\sim 400$~TeV and 1~PeV.
This dip is more pronounced at larger $m_a$, see right panel of
fig.~\ref{fig:IceCubeSpectrum}.  This is in excellent agreement with the
current data, which does not feature any events in this energy range.
Therefore, if this lack of events should become statistically significant in
the future, the boosted DM scenario would provide one possible explanation of
it.  Another interesting aspect of our scenario is that, at low energies, where
the flux is dominated by neutrinos, the expected flavor ratio is $(1:1:1)_E$
after propagation for most decay modes of $a$. Thus the ratio of shower and
track events is predicted to be the same as for the canonical astrophysical
neutrino interpretation at $E_\text{dep} \lesssim \text{few} \times 100$~TeV.
On the other hand, at $E_\text{dep} \sim 1$~PeV, the predicted event rate is
entirely dominated by the DM contribution, which only provides shower events.
This is a unique feature of this model and can be tested with future data.

Let us also remark that a recent IceCube analysis~\cite{Aartsen:2014muf} which
separates events from the northern sky and from the southern sky, exhibits a
noticeable, but not yet statistically significant, bump at energy deposits
around 80~TeV in the southern sky. If this bump should become significant in
the future, it could be interpreted as being due to a relatively large
secondary neutrino flux in the boosted DM scenario.  Since the galactic center,
from where most of these secondary neutrinos are expected to come, is located
in the southern sky, and because neutrinos from the northern hemisphere suffer
some attenuation in the Earth, our model could explain why a similar bump is not observed
in the northern sky.

Note that, without the neutrinos from the 3-body decay $\phi \to \chi \bar\chi
+ (a \to b \bar{b})$, the IceCube fit of our boosted DM scenario becomes much
worse because the prediction would fall short of the observed number of events
at energies $\sim 100$~TeV.  This could be avoided if a mediator with scalar
rather than pseudoscalar couplings to fermions, or a vector boson mediator is
considered.  In this case, the boosted DM scattering cross section would not be
proportional to $(Q^2)^2$, and scattering of $\chi$ particles could explain the
IceCube event excess across the spectrum. However, as we will argue in
sec.~\ref{sec:dd}, direct detection constraints in this case may be
prohibitive.  Ways to avoid these constraints include models with inelastic DM
scattering or with a very small $m_\chi \lesssim 3$~GeV, below the direct detection
threshold.  The second possibility would preclude a simultaneous
explanation of the IceCube events and the galactic center gamma ray excess.

Let us finally discuss the morphology of the IceCube signal from boosted DM.
While the extragalactic flux $d\Phi_\chi^\text{EG} / (dE_\chi \, d\Omega_\psi)$ is
isotropic, the galactic
component $d\Phi_\chi^\text{GC} / (dE_\chi \, d\Omega_\psi)$ peaks
in the galactic center region. (Here $\psi$ denotes the direction of sight.) The
angular resolution in IceCube is about 10$^\circ$--20$^\circ$ for shower
events~\cite{Aartsen:2013jdh}.  With this resolution and more statistics, a
morphology study of the high energy events would provide an important
consistency check of the boosted DM hypothesis.

\section{Dark matter relic density}
\label{sec:relic-density}

An important problem of the boosted DM scenario which we have not addressed yet
is how a particle with a mass of order PeV can account for the observed DM
density in the Universe.  For instance, thermal freeze-out is not a possibility
at masses above few hundred TeV due to unitarity constraints
\cite{PhysRevLett.64.615}.  A long-lived dark matter particle with a mass of
$\mathcal{O}(\text{PeV})$ can nevertheless have the correct abundance in the
Universe~\cite{Moroi:1994rs, Kawasaki:1995cy, Moroi:1999zb, Harigaya:2014waa,
Daikoku:2015vsa}.

Non-thermal production mechanisms for PeV DM include~\cite{Harigaya:2014waa}:
(1) production in cascade decays of the inflaton. In this mechanism, the DM
abundance depends on the number density of inflatons and on the branching ratio
of inflaton decay to DM. (2) production through inelastic scattering between
high energy particles from inflaton decay and the hot plasma. When high-energy
daughter particles scatter on the thermalized plasma, DM can be produced until
the daughter particles' energy become less than $E_\text{th} =
m_\phi^2/(4T)$. (3) For low reheating temperature, DM could be thermally
produced with the correct relic abundance even when the maximum temperature of
the Universe during reheating, $T_\text{max}$, is larger than $m_{\phi}$, as
long as the reheating temperature (defined as the temperature at which the
inflaton energy density equals the radiation energy density) is smaller than
$m_\phi$. The reason is that the continuing decays of the inflaton produce
entropy after DM freeze-out, diluting the DM abundance.  The authors of
ref.~\cite{Harigaya:2014waa} show that these mechanisms can account for the
abundance of DM with $\mathcal{O}$(PeV) mass. Mechanism (2) can achieve this
even if the inflaton does not decay to DM and is thus highly model independent.
PeV DM $\phi$ produced through this mechanism can for instance account for the
observed abundance of DM in the Universe if the reheating temperature of order
10~GeV and the mass of inflaton is of order
$10^{15}$~GeV.~\cite{Harigaya:2014waa}.

\begin{figure*}
  \begin{tabular}{c@{\qquad\qquad}c}
    \includegraphics[height=3.5cm]{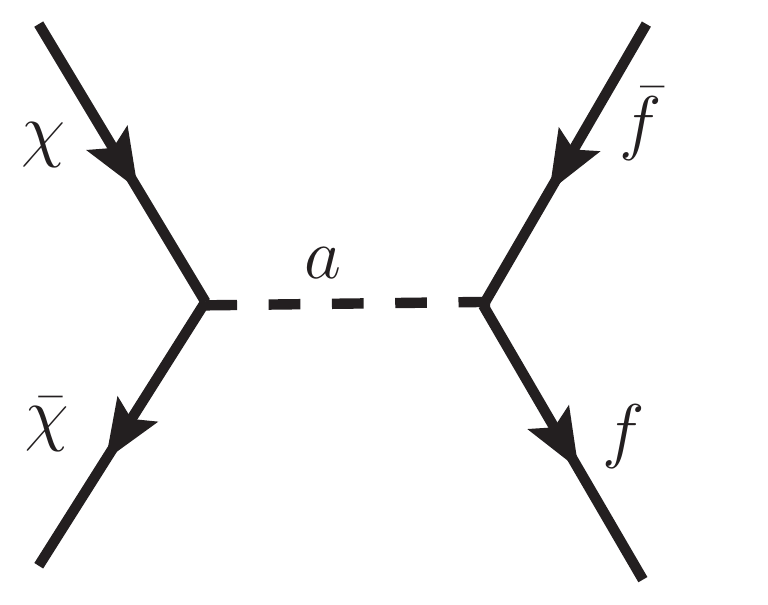} &
    \includegraphics[height=3.5cm]{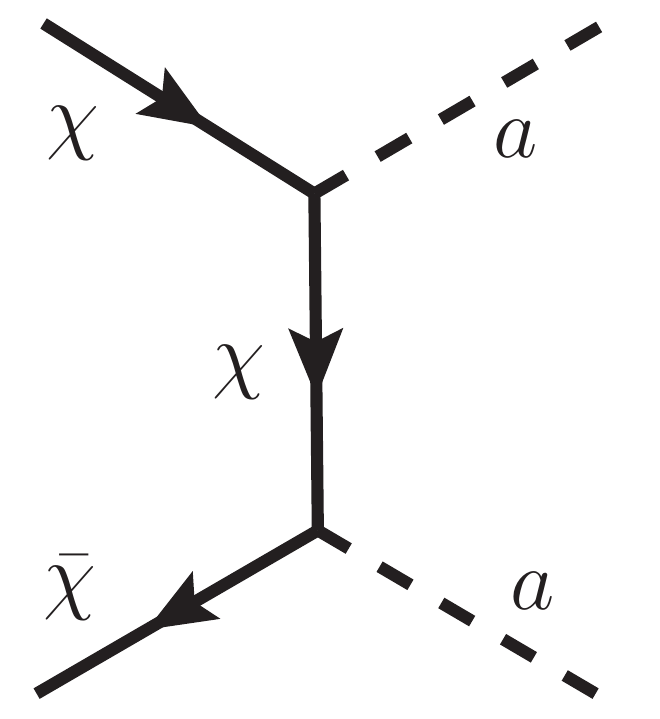} \\
    (a) & (b)
  \end{tabular}
  \caption{The Feynman diagrams for annihilation of the light DM particle $\chi$
    into (a) SM fermions and (b) light pseudoscalar mediator particles $a$.
    (The second process is only possible if $m_a < m_{\chi}$.)}
  \label{fig:ann-feyn}
\end{figure*}

In addition to the non-thermally produced relic abundance of heavy DM particles
$\phi$, there could also be a thermally produced population of the light DM
species $\chi$ if the thermally averaged cross section $\ev{\sigma v_\text{rel}}$ for
$\chi \bar{\chi}$ annihilation through $s$-channel exchange of the mediator $a$
is not too large.  This is naturally realized in our scenario.
$\ev{\sigma v_\text{rel}}$ receives contributions from two classes of
processes, shown in fig.~\ref{fig:ann-feyn}: annihilation to $f \bar{f}$ and,
if $m_a < m_{\chi}$, also annihilation to $aa$.  The thermally averaged
annihilation cross sections read~\cite{Dolan:2014ska}
\begin{align}
  \ev{\sigma v_\text{rel}}_{f \bar{f}} &\simeq
    \sum_f \frac{N_c^f}{2 \pi} \frac{2 g_\chi^2 g_{Y_f}^2 m_\chi^2 m_f^2 / v^2}
                                    {(4m_\chi^2 - m_a^2)^2 + m_a^2 \Gamma_a^2}
    \sqrt{1 - m_f^2 / m_\chi^2} \,,        \label{eq:annihilation-qq} \\
  \ev{\sigma v_\text{rel}}_{aa} &\simeq
    \frac{g_\chi^4 m_\chi}{24 \pi} \frac{(m_\chi^2 - m_a^2)^{5/2}}{(2m_\chi^2 - m_a^2)^4}
    \frac{6 T}{m_\chi} \,,                 \label{eq:annihilation-aa}
\end{align}
where $m_f$ are the SM fermion masses, the sum runs over all SM fermions $f$,
$\Gamma_a$ is the total decay width of $a$, the color factor $N_c^f$ is 3 if $f$ is a quark
and 1 if $f$ is a lepton, and $T$ is the temperature.  The thermally
averaged cross section for annihilation to leptons is completely analogous to
eq.~\eqref{eq:annihilation-qq} except for the color factor. Note that
eqs.~\eqref{eq:annihilation-qq} and \eqref{eq:annihilation-aa} are approximate
results, with only the leading terms in the relative velocity $v_\text{rel}$ kept. The
proportionality to $T$ in eq.~\eqref{eq:annihilation-aa} arises because the
process $\chi \bar{\chi} \to a a$ is $p$-wave suppressed. When evaluating
$\ev{\sigma v_\text{rel}}_{aa}$ for calculating the relic density of $\chi$, we set $T$ to
its typical value at freeze-out: $T_F \simeq m_\chi /
20$~\cite{Agashe:2014kda}. Due to the temperature dependence, annihilation to
$aa$ can be important in determining the thermal relic abundance of $\chi$, but
does not lead to observable indirect signals today, where the relic population
of $\chi$ is non-relativistic. $\chi \bar{\chi} \to f \bar{f}$, on the other
hand, is an $s$-wave process and is therefore relevant both today and in the
early Universe.

At our first benchmark point from table~\ref{tab:benchmarks} ($m_a = 12$~GeV),
it is indeed the interplay of the annihilation processes $\chi \bar\chi \to a
a$ and $\chi \bar\chi \to b \bar{b}$ that sets the relic density of $\chi$,
$f_\chi \simeq 0.6$.  At the second benchmark point ($m_a = 80$~GeV),
annihilation to $a a$ is kinematically forbidden at freeze-out, therefore $\chi
\bar\chi \to b \bar{b}$ accounts for the relic density $f_\chi \simeq 0.33$
alone.

In fact, the thermal production of $\chi$ has some subtlety to it if the
abundance of the heavy species $\phi$ is explained by a low reheating
temperature $T_\text{RH}$. The freeze-out temperature $T_F$ of $\chi$ is of
order $T_F \sim m_\chi / 20 \sim 1.5$~GeV at our benchmark points. If
$T_\text{RH} \lesssim T_F$, the relic abundance $\Omega_\chi$ of $\chi$ will be
smaller than predicted from the naive estimate for Dirac fermions, $\Omega_\chi h^2 \sim
6 \times 10^{27}\ \text{cm$^3$ sec$^{-1}$} / \ev{\sigma v_\text{rel}}$.  If $T_\text{RH}
\gg T_F$, the thermal production of $\chi$ is not affected. This is possible
with a $\sim 10^{15}$~GeV inflaton field with $T_\text{RH} \sim 10$~GeV that
could provide the correct relic abundance for $\phi$~\cite{Harigaya:2014waa}.
For simplicity, we assume in the following that this second case is realized.
We moreover assume in the following that $\phi$ and $\chi$ have comparable
relic density, and that together they account for all the DM in the Universe
(i.e.\ $f_\phi + f_\chi = 1$).

\section{The galactic center gamma ray excess}
\label{sec:GCE}

The fact that the light DM species $\chi$ in the boosted DM scenario
can have a non-negligible relic abundance
and a relatively large annihilation cross section to SM fermions in the present day
Universe indicates that there may be interesting indirect signatures,
in addition to the primary signal from highly boosted
$\chi$ particles from $\phi$ decay.

In particular, the boosted DM scenario can fit the excess of gamma rays which
has been observed from the direction of the galactic center at energies of few
GeV~\cite{Goodenough:2009gk, Hooper:2010mq, Daylan:2014rsa}. It has been argued
that, if the dominant DM annihilation channel is $\chi \bar\chi \to b \bar{b}$,
as in our boosted DM scenario, a 30--40~GeV DM particle with $\ev{\sigma
v_\text{rel}}_{b \bar{b}}$ in the range $1.4$--$2.0 \times 10^{-26}
\text{cm$^3$ sec$^{-1}$}$ provides a good fit to the data.
Since in our scenario the light DM species $\chi$ constitutes only a fraction
$f_\chi$ of the total DM relic density, its annihilation cross section today has
to be correspondingly larger by $1/f_\chi^2$.

At our benchmark points from table~\ref{tab:benchmarks}, the predicted annihilation
cross sections are $\ev{\sigma v_\text{rel}}_{b \bar{b}} \sim 2.8 \times 10^{-26}\
(18 \times 10^{-26})\ \text{cm$^3$/sec}$.
Here the first
number stands for the benchmark point with $m_a = 12$~GeV, while the second
one (in parenthesis) is for the benchmark point with $m_a = 80$~GeV.
With $f_\chi = 0.6\ (0.33)$ (see sec.~\ref{sec:relic-density}), and taking
into account that we chose $m_\chi \sim 30$~GeV at the benchmark points, we thus
see that the galactic center gamma ray excess could be explained by our
boosted DM scenario.  Note that for the special case $m_a \sim 2 m_\chi$,
this could be achieved even for much smaller couplings $g_{Y_b}$ and $g_\chi$
because the annihilation would be resonantly enhanced.

\section{Constraints}
\label{sec:constraints}

Constraints on the boosted DM scenario arise on the one hand from indirect DM
searches sensitive to high-energy particles from the 3-body decay $\phi
\to \chi \bar{\chi} a$, followed by decay of the mediator $a$ into SM particles
including positrons and gamma rays. We will discuss these possibilities in
secs.~\ref{sec:positron} and \ref{sec:gamma}, respectively.  On the other hand,
direct DM searches could hope to directly observe the relic population of light
DM particles $\chi$, see sec.~\ref{sec:dd}.  Finally, the mediator $a$ could be
directly produced in accelerator experiments, leading to constraints as well
(see sec.~\ref{sec:flavor-collider}).

\subsection{Positron flux from 3-body decay $\phi \to \chi \bar{\chi} a$}
\label{sec:positron}

The $e^\pm$ flux at any given point $\vec{x}$ in the galaxy is given
by~\cite{Cirelli:2010xx}
\begin{align}
  \frac{d\Phi_{e^\pm}(E_e,\vec{x})}{dE_e} &=
    \frac{1}{b(E_e,\vec{x})} \,
    \frac{\rho(\vec{x})}{m_\phi} \,
    \Gamma_3(\phi \to \chi \bar\chi a) \sum_f \BR(a \to f \bar{f})
    \int^{m_\phi/2}_{E_e} \! dE_e^S \, \frac{dN^f_{e^\pm}(E_e^S)}{dE_e^S}
      I(E_e,E_e^S,\vec{x}) \,,
  \label{eq:positron-spectrum}
\end{align}
where $\rho(\vec{x})$ gives the DM density distribution in the galaxy,
$\Gamma_3(\phi \to \chi \bar\chi a)$ is the 3-body decay rate from
eq.~\eqref{eq:Gamma-3body}, $E_e^S$ is
the $e^\pm$ energy at production, and $dN^f_{e^\pm}(E_e^S) / dE_e^S$ is the $e^\pm$
spectrum at production for $a$ decay to $f \bar{f}$. We obtain
$dN^f_{e^\pm}(E_e^S) / dE_e^S$ in analogy to the secondary neutrino spectrum
discussed in sec.~\ref{sec:icecube-neutrino} by folding the $e^\pm$ spectrum in the $a$
rest frame (taken from~\cite{Cirelli:2010xx}) with the energy distribution of $a$
particles from $\phi \to \chi \bar{\chi} a$ (see eq.~\eqref{eq:Gamma-3body} and
fig.~\ref{fig:Eadistribution}).
The sum in eq.~\eqref{eq:positron-spectrum} runs over all final
states of $a$ decay, and $\BR(a \to f \bar{f})$ are the corresponding
branching ratios. The factor $b(E_e,\vec{x})$ describes energy loss during
propagation~\cite{Cirelli:2010xx}. Finally, $I(E_e,E_e^S,\vec{x})$ is the
generalized halo function, which can be understood as a Green's function of the
diffusion-loss equation, describing the probability for an $e^\pm$ with initial
energy $E_e^S$ to be detected with energy $E_e$.  We take the halo function from
ref.~\cite{Cirelli:2010xx}, assuming a Navarro-Frenk-White (NFW) DM density
profile~\cite{Navarro:1995iw} and the \textit{MED} propagation model
\cite{Delahaye:2007fr}.  The dependence of our results on the DM density profile is
quite small because the dark matter decay rate only depends linearly on the DM
density.  The uncertainty from the propagation model could change our
constraints, but we have checked that even for the propagation model
\textit{MAX} from ref.~\cite{Delahaye:2007fr}, the predicted flux is at most
a factor of 2 larger than for the \textit{MED} model.

\begin{figure*}
  \begin{tabular}{cc}
    \includegraphics[width=0.45\textwidth]
      {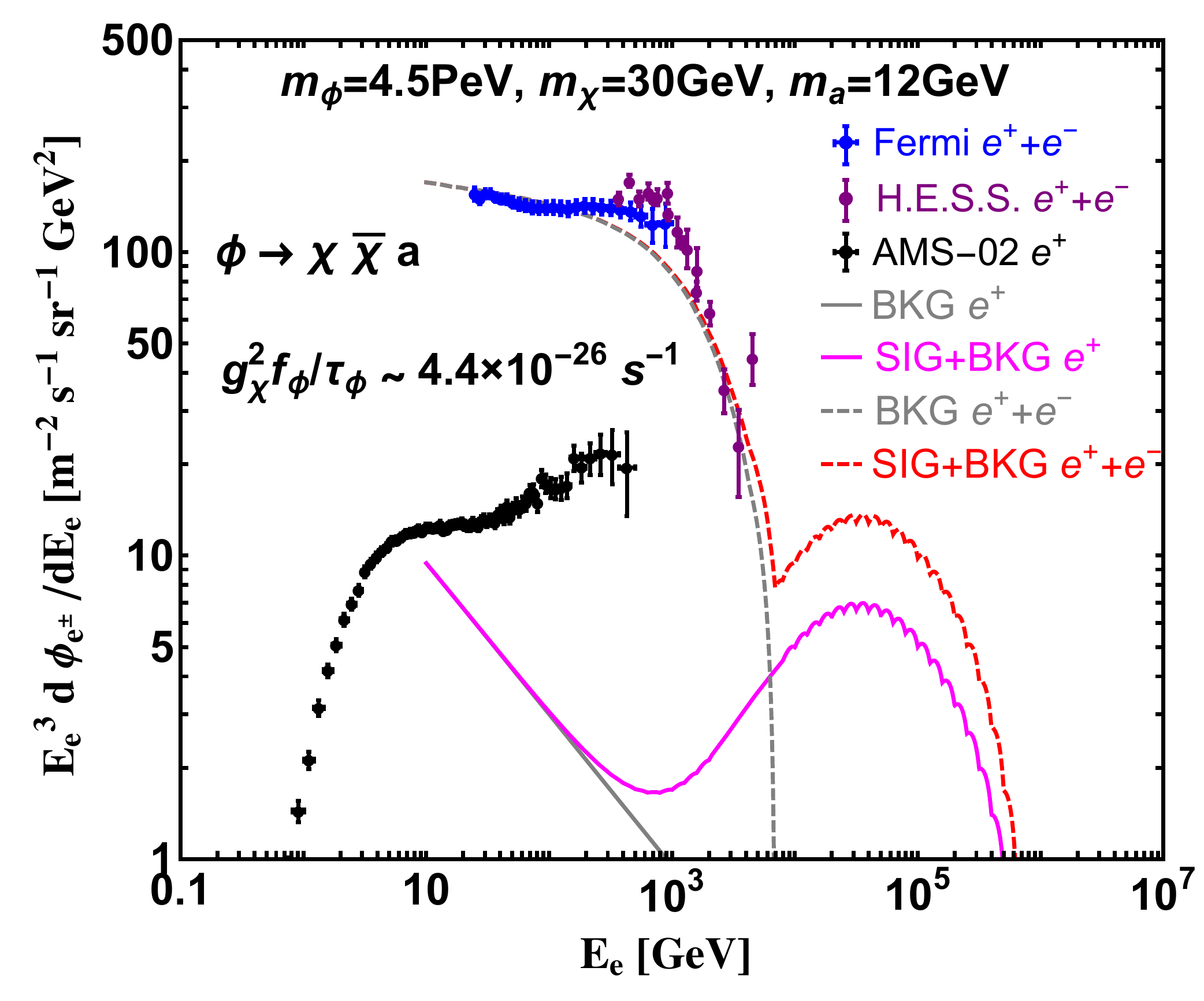} &
    \includegraphics[width=0.45\textwidth]
      {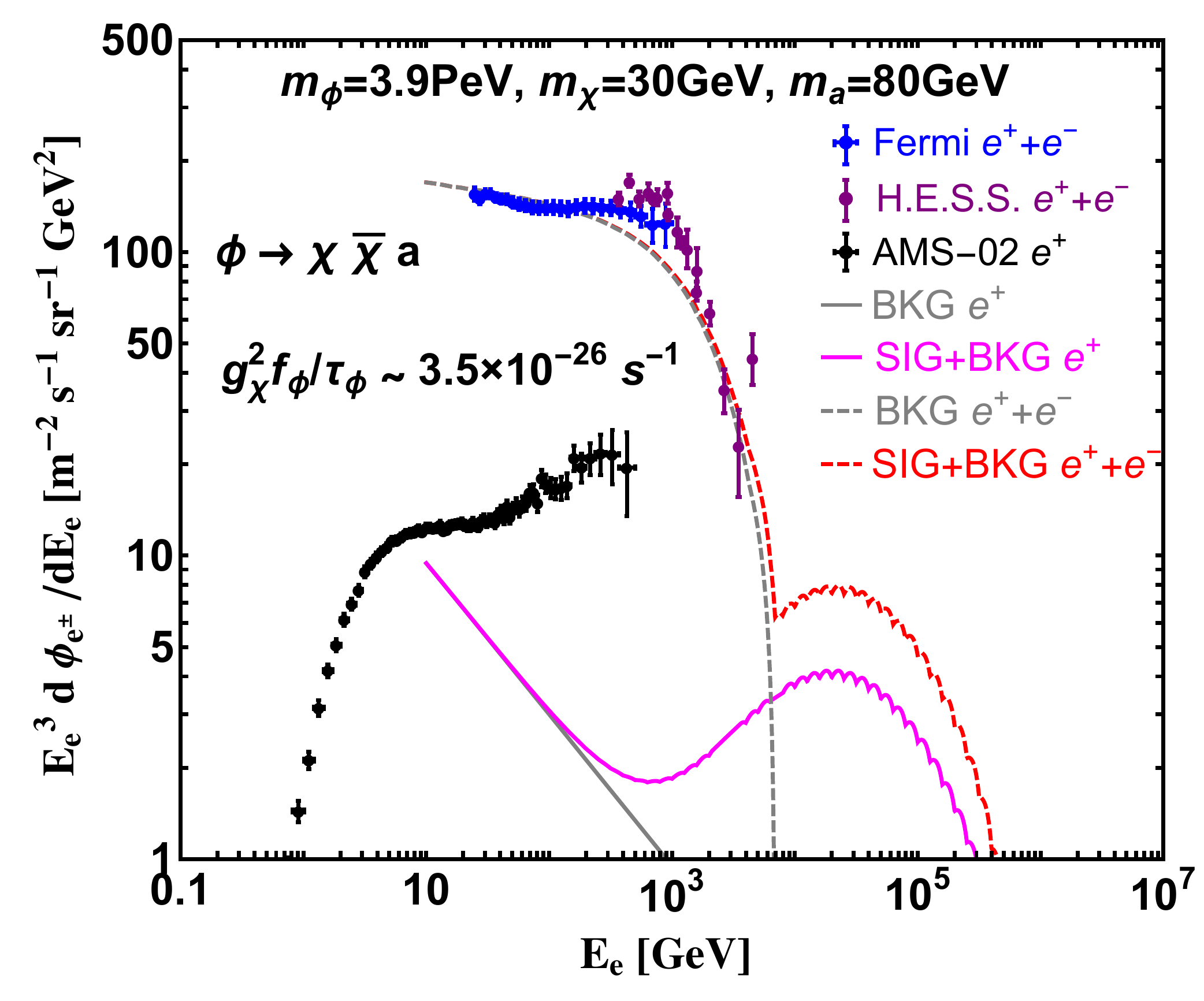} \\
    (a) & (b)
  \end{tabular}
  \caption{The positron flux from $\phi \to \chi \bar{\chi} a$ decay,
    where $a$ decays dominantly to $b \bar{b}$. The parameters in the
    \textit{left panel} (\textit{right panel}) are fixed at
    $m_\phi = 4.5$~PeV (3.9~PeV) for the heavy DM mass, $m_\chi = 30$~GeV for the light
    DM mass, and $m_a = 12$~GeV (80~GeV) for the mediator mass.
    The AMS-02 positron flux data \cite{Aguilar:2014mma}, as well as
    the Fermi-LAT \cite{Ackermann:2010ij} and H.E.S.S.
    \cite{Aharonian:2008aa, Aharonian:2009ah} data for the combined
    electron plus positron flux are plotted as well.}
  \label{fig:posiflux}
\end{figure*}

In fig.~\ref{fig:posiflux}, we have plotted the positron flux at Earth from the
$\phi \to \chi \bar\chi a$ decay, where $a$ dominantly decays into $b \bar{b}$.
We fix the mass parameters at our benchmark values $m_\phi = 4.5$~PeV (3.9 PeV),
$m_\chi = 30$~GeV and $m_a = 12$~GeV (80~GeV) in the left panel (right panel).
Once the masses are fixed, $d\Phi_{e^\pm}/dE_e$
depends on the model parameters through the ratio $g_\chi^2 f_\phi / \tau_\phi$.

The background model for the $e^+$ flux is taken from
refs.~\cite{Baltz:1998xv, Moskalenko:1997gh}, while the background model for
the combined $e^+ + e^-$ flux is taken as a fitting function from
ref.~\cite{Aharonian:2008aa}.
We compare to the AMS-02 $e^+$ flux data
\cite{Aguilar:2014mma} as well as the Fermi-LAT \cite{Ackermann:2010ij} and
H.E.S.S. \cite{Aharonian:2008aa, Aharonian:2009ah} $e^+ + e^-$ flux data
to provide a constraint on this decay. Note that when comparing to Fermi-LAT
and H.E.S.S. data, which includes both $e^+$ and $e^-$, the signal flux is twice the
$e^+$ signal flux. The error bars in the H.E.S.S. data do not contain systematic
uncertainties, while those in the Fermi-LAT and AMS-02 data do. By requiring
that the signal flux should be outside the $1\sigma$ error bar for any of these
data points, we find constraints on the coupling $g_\chi$, the relative abundance
of the heavy DM $f_\phi$, and its lifetime $\tau_\phi$:
\begin{align}
  \begin{split}
    \frac{g_\chi^2 f_\phi}{\tau_\phi} &\lesssim
      4.4 \times 10^{-26}\ \text{sec$^{-1}$} \qquad \text{for $m_a = 12$~GeV} \,, \\
    \frac{g_\chi^2 f_\phi}{\tau_\phi} &\lesssim
      3.5 \times 10^{-26}\ \text{sec$^{-1}$} \qquad \text{for $m_a = 80$~GeV} \,.
  \end{split}
  \label{eq:econstraint}
\end{align}
We see from table~\ref{tab:benchmarks} that our two benchmark points
easily satisfy these constraints.

The cosmic electron background is complicated and model dependent. The
background model from ref.~\cite{Aharonian:2008aa} has a lot of parametric
freedom regarding in particular the overall normalization, which could alleviate
the constraints.
Our constraints should therefore be considered as very conservative. Even for
the most conservative assumption of zero background, we would still obtain a
constraint on $g_\chi^2 f_\phi / \tau_\phi$ by requiring that the predicted signal
does not significantly overshoot the data. The dominant constraint in this case
would come from the last two bins of H.E.S.S. data, and the constraint would be
weaker by a factor of $\sim 5$ compared to eq.~\eqref{eq:econstraint}. Also
including the systematic error of the H.E.S.S. data would make the constraint
even weaker.

\subsection{Gamma ray flux from 3-body decay $\phi \to \chi \bar{\chi} a$}
\label{sec:gamma}

The secondary gamma ray flux from the decay $\phi \to \chi \bar{\chi} a$ may
contribute to gamma ray searches, in particular to gamma ray searches in the
galactic center region and in measurements of the diffuse isotropic gamma ray
flux, i.e.\ the residual flux obtained after subtracting the contribution from
known astrophysical sources.  We focus here on the diffuse flux because we will
see that the strongest limits are coming from air shower detectors located in
the northern hemisphere and thus unable to observe the galactic
center~\cite{Ahlers:2013xia}.  The only exception is a $\gamma$ ray search
carried out by the IceCube collaboration using the IceTop
array~\cite{Aartsen:2012gka}. This search, however, is only sensitive at energies
above 1~PeV, where the secondary $\gamma$ ray flux from decay of $\sim 4$~PeV
DM particles is already negligible.
Moreover, it is worth emphasizing that searching for signals of decaying
DM in DM-rich, but also foreground-rich, regions like the galactic center is
much less promising than searching for annihilating DM in these regions. The
reason is that the DM decay rate depends linearly on the DM density
$\rho(\vec{x})$, while the annihilation rate scales as $\rho(\vec{x})^2$.

The procedure for calculating the diffuse gamma ray flux is
similar to the one for the secondary neutrino fluxes described
in sec.~\ref{sec:ic} and for the $e^\pm$ fluxes described in sec.~\ref{sec:positron}.
In particular, we can use eqs.~\eqref{eq:DMflux0} and
\eqref{eq:DMflux2} after replacing $E_\chi$ by the $\gamma$ energy $E_\gamma$
and the DM spectrum $dN_\chi / dE_\chi$ by the gamma ray spectrum at production
$dN_\gamma / dE_\gamma$.  Note that $dN_\gamma / dE_\gamma$ must be normalized
such that its integral over $E_\gamma$ gives the average number of photons produced
in each $a$ decay, accounting for two body decays without photon emission and
for three body decays that lead to the radiation of photons.
We obtain $dN_\gamma / dE_\gamma$ by boosting the $\gamma$ ray
spectra in the $a$ rest frame (taken from~\cite{Cirelli:2010xx}) into the lab
frame according to the energy spectrum of $a$ particles given by
eq.~\eqref{eq:Gamma-3body} and fig.~\ref{fig:Eadistribution} and
multiplying by $\BR_3(\phi \to \chi \bar\chi a)$
from eq.~\eqref{eq:br3}.  For the gamma ray flux, also an absorption factor of
the form
\begin{align}
  \exp[-\text{Abs}(E_\chi, z)]
  \label{eq:attenuation}
\end{align}
must be included in eq.~\eqref{eq:DMflux2} to describe the attenuation of
extragalactic gamma rays on their way from the source to us. We take this
factor from ref.~\cite{Cirelli:2010xx}.

We then obtain the diffuse gamma ray flux conservatively according to the
formula~\cite{Cirelli:2012ut}
\begin{align}
  \frac{d\Phi_\text{diffuse}}{dE_\gamma}
    = \frac{d\Phi_\text{EG}}{dE_\gamma}
    + 4\pi \frac{d\Phi_\text{GC}}{dE_\gamma \, d\Omega} \bigg|_\text{minimum} \,.
  \label{eq:diffuse-flux}
\end{align}
Here, $d\Phi_\text{GC} / (dE_\gamma \, d\Omega)|_\text{minimum}$ denotes
the minimum of the differential galactic flux over solid angles, which we take to be the
flux from the direction opposite to the galactic center~\cite{Cirelli:2012ut}.
We have checked that using instead the average of the differential flux over
a cone with opening angle $90^\circ$, centered around the direction opposite to
the galactic center, would change $d\Phi_\text{GC} / (dE_\gamma \, d\Omega)
|_\text{minimum}$ by $\mathcal{O}(20\%)$.

\begin{figure*}
  \begin{tabular}{cc}
    \includegraphics[width=0.45\textwidth]
      {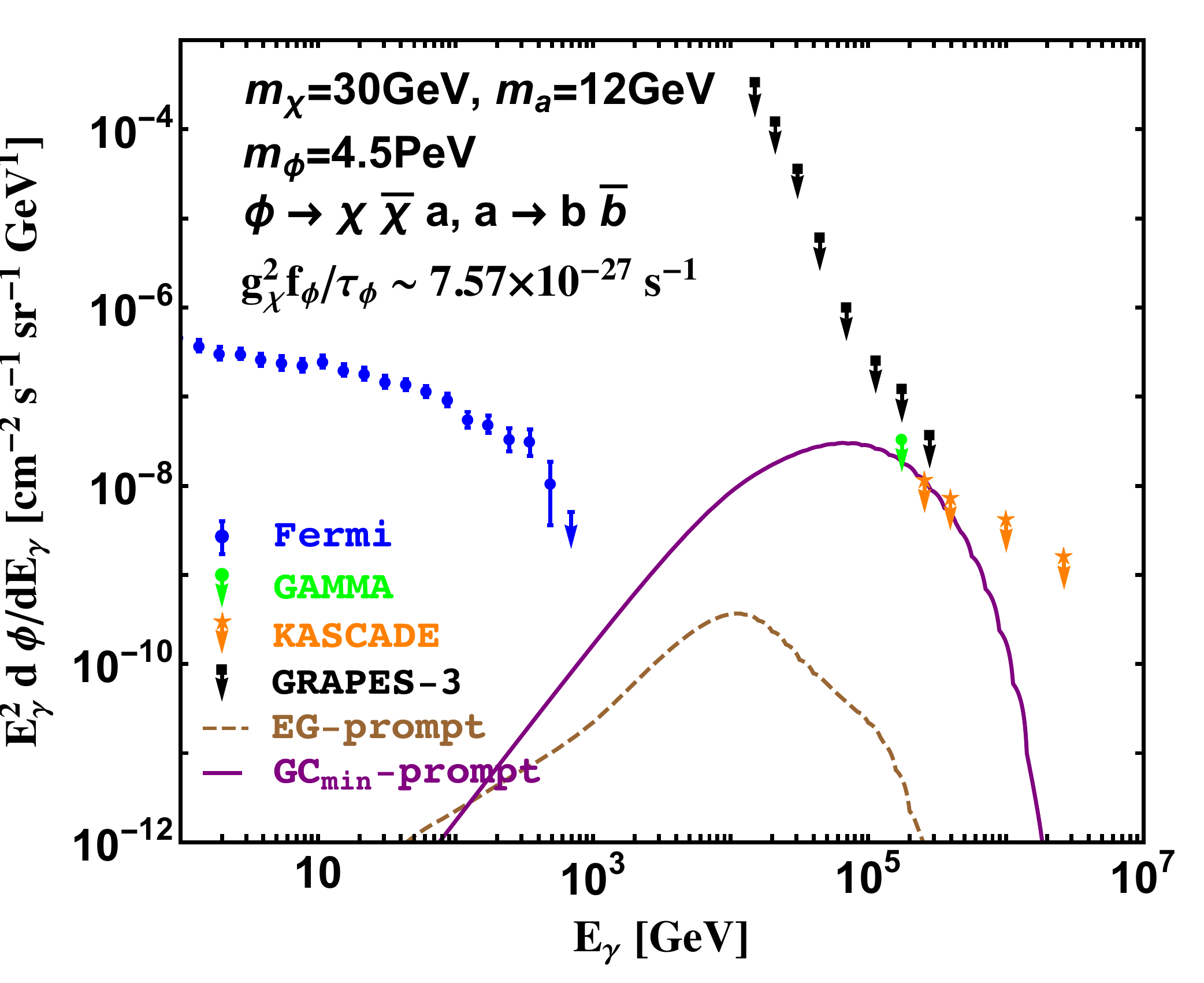} &
    \includegraphics[width=0.45\textwidth]
      {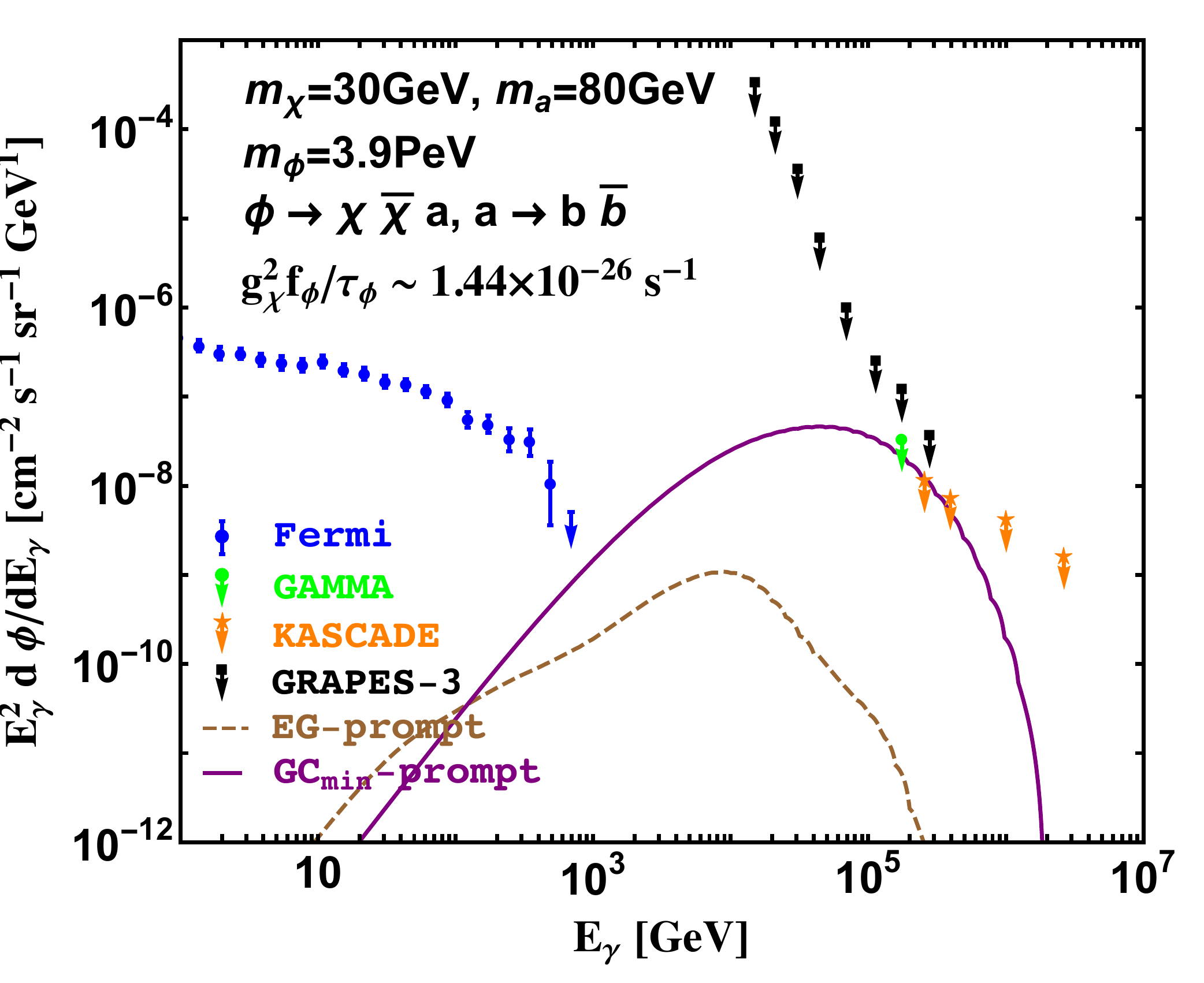} \\
    (a) & (b)
  \end{tabular}
  \caption{The diffuse galactic (solid purple) and extragalactic (dashed brown)
    gamma ray fluxes
    from $\phi \to \chi \bar\chi a$ decay, followed by $a \to \bar bb$.  The
    galactic flux is assumed to have in every direction the magnitude it has in
    the direction opposite to the galactic center~\cite{Cirelli:2012ut},
    evaluated assuming a Navarro-Frenk-White DM density profile~\cite{Navarro:1995iw}.
    We include only prompt gamma rays,
    neglecting the low energy contribution from inverse Compton scattering because
    we have checked that the limit is dominated by the prompt signal.
    We compare to the Fermi-LAT
    measurement of the diffuse gamma ray flux from
    ref.~\cite{Ackermann:2014usa}, using foreground model~C defined in this
    reference, and to the limits from air shower detectors~\cite{Schatz:2003aw,
    Minamino:2009ds, Martirosov:2009ni}.  The
    model parameters are fixed at the values given by our first (second) benchmark
    point from table~\ref{tab:benchmarks} in the \textit{left panel}
    (\textit{right panel}).}
  \label{fig:gammaflux}
\end{figure*}

We plot the galactic and extragalactic contributions to the diffuse gamma ray
flux in fig.~\ref{fig:gammaflux}.
We see that the contribution from $\phi$ decay in the galaxy dominates over
the extragalactic flux due to the attenuation factor eq.~\eqref{eq:attenuation},
which suppresses the extragalactic gamma ray flux.

Note that we neglect the low energy contribution from inverse Compton
scattering (ICS) of high-energy $e^\pm$ from the decay of heavy DM $\phi$ on CMB photons,
starlight, and light rescattered on dust.  We estimate~\cite{Profumo:2009uf,
Longair:1992} that the energy spectrum of ICS photons induced by $\phi$ decay
peaks at 1--100~GeV.  Following~\cite{Esmaili:2014rma}, we have then estimated that
the energy density in ICS gamma rays predicted at our benchmark points
is at least one order of magnitude lower than the energy density
measured by Fermi-LAT at 1--100~GeV~\cite{Esmaili:2014rma,
Ackermann:2014usa}. Similarly,
also the contribution from bremsstrahlung of $e^\pm$ on dust is negligible.

To set limits on the parameter space of boosted DM, we compare to the
diffuse gamma ray spectra from Fermi-LAT~\cite{Ackermann:2014usa}
and to the flux limits from the air shower detectors KASCADE~\cite{Schatz:2003aw},
GRAPES-3~\cite{Minamino:2009ds} and GAMMA~\cite{Martirosov:2009ni}, see
also~\cite{Ahlers:2013xia}. From fig.~\ref{fig:gammaflux},
we see that the constraint will come mostly from the air shower detectors
and the last bin of Fermi-LAT data.  By requiring that the
predicted signal is smaller than the limit from the air shower detectors,
we obtain the constraints
\begin{align}
  \begin{split}
    \frac{g_\chi^2 f_\phi}{\tau_\phi} &\lesssim
      0.76 \times 10^{-26}\ \text{sec$^{-1}$} \qquad \text{for $m_a = 12$~GeV} \,, \\
    \frac{g_\chi^2 f_\phi}{\tau_\phi} &\lesssim
      1.44 \times 10^{-26}\ \text{sec$^{-1}$} \qquad \text{for $m_a = 80$~GeV} \,.
  \end{split}
  \label{eq:gamma-constraint}
\end{align}
We see that both of our benchmark points from table~\ref{tab:benchmarks}
satisfy these constraint.

\subsection{Direct detection}
\label{sec:dd}

In the boosted DM scenario, conventional DM direct detection experiments can
only constrain the thermally produced population of light DM particles $\chi$,
not the population of heavy DM particles $\phi$. The density of $\phi$
particles and thus also the flux of boosted $\chi$ particles from $\phi$ decay
are too small to be observed in these detectors. Therefore our discussion of
direct detection will focus on the non-relativistic population of the light DM
species $\chi$.  The cross section for $\chi$--nucleus scattering
is~\cite{Dolan:2014ska}
\begin{align}
  \frac{d\sigma}{dE_r}
    = \frac{m_T}{32\pi} \frac{1}{v^2} \frac{g_\chi^2}{(Q^2 + m_a^2)^2}
      \frac{(Q^2)^2}{m_N^2 m_\chi^2}
      \sum_{N,N' = p,n} g_N g_{N'} F_{\Sigma''}^{N,N'}  \,,
\end{align}
where $E_r$ is the nuclear recoil energy, $v$ is the DM velocity, $Q^2 = 2 m_T
E_r \sim 100\ \text{MeV}^2$ is the 4-momentum transfer squared, $m_T$ is the mass of the
target nucleus and $m_N$ is the nucleon mass.  The quantities $F_{\Sigma''}^{N,N'}$
are the pseudoscalar form factors of the target nucleus (see e.g.\
\cite{Fitzpatrick:2012ix}), and the effective nucleon couplings $g_N$, $g_N'$
depend on the $g_{Y_f}$ (see also ref.~\cite{Arina:2014yna}).
For our choice $m_a \gtrsim 10$~GeV, we have $m_a^2 \gg Q^2$, so that $Q^2$ is
negligible in the denominator.  The factor $(Q^2)^2$ in the numerator arises
because, in the non-relativistic limit, $\bar\chi \gamma^5 \chi \propto
\sqrt{Q^2}$.  Direct detection constraints are in general very weak in our
boosted DM model due to the $(Q^2)^2$ suppression unless the mediator mass $m_a$ is
extremely small.  The resulting limit on $g_\chi g_{Y_f}$ is therefore much weaker than the
value needed by the thermal relic density~\cite{Dolan:2014ska}.

Departing for a moment from our toy model with a pseudoscalar mediator, we note
that in general, boosted DM models with interaction cross sections strong
enough to explain the IceCube events would also lead to a large signal in
direct detection experiments.  From a model building point of view, there are
several ways of circumventing this, other than using a pseudoscalar coupling as
in our toy model.  (1) Construct a model in which the scattering of the light
DM particles $\chi$ on nuclei is inelastic \cite{TuckerSmith:2001hy}. If the
mass splitting $\delta m$ between the ground state of $\chi$ and the excited
state $\chi^*$ which is produced in the scattering is sufficiently large, it
will lead to vanishing event rates in direct searches, but will have no
influence on boosted DM collisions as long as $\delta m$ is small compared to
the energy of the boosted DM particles.  (2) Assume the relic abundance of the
light DM species is sufficiently low to avoid direct detection limits. This
would of course preclude a simultaneous explanation of the IceCube events and
the galactic center gamma ray excess. (3) Choose the light DM mass smaller than
$\sim 3$~GeV, below the energy threshold for direct detection. This would also
preclude an explanation of the galactic center gamma ray excess.

\subsection{Constraints from flavor physics experiments and from collider searches}
\label{sec:flavor-collider}

In the following, we discuss constraints on our boosted DM scenario from
experiments at flavor factories and at high energy colliders and indicate for
each constraint to which of the three renormalizable models from
sec.~\ref{sec:models} it applies.

A large number of constraints arises from Kaon and $B$ meson decays
\cite{Dolan:2014ska}. Searches are sensitive to the production of the
pseudoscalar $a$ in decays of these mesons if $a$ subsequently decays to
leptons, photons or invisible particles. Since we are considering the case $m_a
\gtrsim 10$~GeV, those constraints are, however, significantly weakened by the
fact that $a$ would have to be off-shell.

$B_s \to \mu^+ \mu^-$ is the only search channel sensitive to an off-shell
pseudoscalar.  If we consider a renormalizable model for the pseudoscalar $a$
in the framework of a Two Higgs Doublet Model, as in the \textit{MSSM-like} and
\textit{Flipped} models from sec.~\ref{sec:models}, $a$ couples to the SM by
mixing with the heavy pseudoscalar $A^0$. The mixing angle is denoted by
$\theta$.  The branching ratio for $B_s \to \mu^+ \mu^-$ in the
\textit{MSSM-like} model is given in ref.~\cite{Skiba:1992mg, Ipek:2014gua}.
The contribution from $a$ to the amplitude is proportional to $\tan^2\beta
\sin^2\theta$.  The constraint for $m_a \sim 10$~GeV is $\tan\beta \sin\theta =
\sqrt{2 g_{Y_d} g_{Y_\mu}} \lesssim 0.4\ (0.51)$ for charged Higgs boson masses
of $m_{H^\pm} \sim 800\ (400)$~GeV, while the constraint for $m_a \sim 80$~GeV
is about $\tan\beta \sin\theta = \sqrt{2 g_{Y_d} g_{Y_\mu}} \lesssim 3.8\
(4.8)$~\cite{Ipek:2014gua}.  For the \textit{Flipped} model, where lepton
couplings are proportional to $\cot\beta$, the amplitude from $a$ exchange is
proportional to $\tan\beta \cot\beta \sin^2 \theta = \sin^2 \theta$. Therefore,
the constraint is $\sin\theta = \sqrt{2 g_{Y_d} g_{Y_\mu}} \lesssim 0.4\
(0.51)$ for charged Higgs boson masses of $m_{H^\pm} \sim 800\ (400)$~GeV when
$m_a \sim 10$~GeV.  For $m_a \sim 80$~GeV, there is no constraint on
$\sin\theta$.  Note that the $B_s \to \mu^+ \mu^-$ constraint does not apply to
the \textit{Vector-quark} model because $a$ does not couple to leptons in this
model.

An additional constraint, which is independent of the couplings of the pseudoscalar
$a$ to fermions, arises from the exotic decay $h \to a a$. In the context of the
\textit{MSSM-like} and \textit{Flipped} models, the branching ratio for
this decay is constrained by~\cite{ATLAS:2013sla, CMS:aya, Ipek:2014gua}
\begin{align}
  \BR(h \to aa) \simeq 0.02\ \bigg( \frac{m_A}{800\ \text{GeV}} \bigg)^4
                             \bigg( \frac{\sin\theta}{0.01}    \bigg)^4 < 0.22 \,,
\end{align}
where $m_A$ is the mass of the heavy pseudoscalar.
If $m_A \simeq 800$~GeV, $\sin \theta$ has to be smaller than $0.02$.
If $m_a$ becomes comparable to $m_h/2$, the above constraint is weakened, and
for $m_a > m_h/2$ it is completely absent. It is also absent in the
\textit{Vector-quark} model.

We should also consider constraints from the LEP experiments, which have
searched for $e^+ e^- \to h A^0$, where $A^0$ is the pseudoscalar Higgs boson
appearing in the MSSM~\cite{Schael:2006cr}.  While these searches exclude
$A^0$ masses below 90~GeV, they do not apply to models with an extra pseudoscalar
$a$, like the scenarios we are considering here~\cite{Kozaczuk:2015bea}.

If $a$ is heavy enough to decay to $\chi \bar\chi$,
ref.~\cite{Izaguirre:2014vva} shows that searches for $b$ jets and missing
energy can provide an excellent constraint on the pseudoscalar $a$. The
dominant processes are $g g \to b \bar{b} a$ and $\parenbar{b} g \to
\parenbar{b} a$, with $a$ decaying to $\chi \bar\chi$ subsequently.  The
current CMS and ATLAS searches \cite{Aad:2013ija, CMS:2014nia}, which are
optimized for final states with two $b$ quarks, lead to the constraint
$\sqrt{g_\chi g_{Y_b}} \lesssim 5$ for $m_a \sim 100$--250~GeV and assuming
$g_\chi = g_{Y_b} \sqrt{2} m_b / v$~\cite{Izaguirre:2014vva}.  If $g_\chi$ is
significantly larger than $g_{Y_b} \sqrt{2} m_b / v$, the limit will become
somewhat weaker since the probability for radiating an on-shell $a$ particle
changes~\cite{Izaguirre:2014vva}.

In the intermediate mass region 20--80~GeV, ref.~\cite{Kozaczuk:2015bea} also
discusses the processes $g g \to b \bar{b} a$ and $\parenbar{b} g \to
\parenbar{b} a$, but considering the subsequent decays $a \to \mu^+ \mu^-$ and
$a \to \tau^+ \tau^-$.  By looking for these leptonic final states, the high
luminosity LHC can be sensitive to $g_{Y_b} \sim 7$ with 100~fb$^{-1}$ of
integrated luminosity, assuming that $g_{Y_f}$ is universal for down type
quarks and charged leptons (as in the \textit{MSSM-like} model).  Since this
assumption is not satisfied in the \textit{Flipped} model, which has suppressed
couplings of $a$ to leptons, and in the \textit{Vector-quark} model, in which $a$
does not couple to leptons at tree level, the constraint would be significantly
weaker or completely absent in these models.

In the light mass region $m_a \sim 5.5$--14~GeV, CMS has searched for $a \to
\mu^+ \mu^-$ in the context of the Next-to-Minimal Supersymmetric Standard
Model \cite{Chatrchyan:2012am}. The upper limit on the cross section for the
process $p p \to a \to \mu^+ \mu^-$ is around 2--4~pb. This translates into a
constraint of $g_{Y_d} \sim 2$ in the \textit{MSSM-like} model, where
$g_{Y_\ell} = g_{Y_d}$~\cite{Kozaczuk:2015bea}.  The \textit{Flipped} and
\textit{Vector-quark} models are not restricted by this constraint due to
the smallness or complete absence of leptonic couplings of $a$.

Let us summarize the most stringent constraints for the three models defined in
sec.~\ref{sec:models} (see also the last column in table~\ref{tab:constraints}
below).  For the \textit{MSSM-like} model, the most stringent limit comes from
$B_s \to \mu^+ \mu^-$. It rules out the \textit{MSSM-like} model as a
UV-completion for our $m_a = 12$~GeV benchmark point, while for the $m_a =
80$~GeV benchmark point, it is a viable possibility.  For the \textit{Flipped}
model, the coupling between leptons and the pseudoscalar $a$ is suppressed once
we are in the large $\tan\beta$ region. But the constraint from $h \to a a$
still implies that the mixing angle $\sin\theta$ between $a$ and the heavy
pseudoscalar $A^0$ should be very small.  If we require that $\tan\beta
\lesssim 50$, this disfavored also the \textit{Flipped} model as a UV completion
for our $m_a = 12$~GeV benchmark point. At $m_a = 80$~GeV, the $h \to aa$
constraint is absent because the decay is kinematically forbidden.  For the
\textit{Vector-quark} model, only the perturbativity of the Yukawa couplings
involving the heavy quarks, together with the LHC limits on their mass, imposes
a very weak constraint $g_{Y_b} \lesssim 20$~\cite{Izaguirre:2014vva}.

\section{Summary and Conclusion}
\label{sec:conclusions}

\newcommand{\mr}[1]{\multirow{3}{*}{#1}}
\begin{table}
  \centering
  \begin{ruledtabular}
  \begin{tabular}{cccc@{}ccccl@{\hspace{-0.3cm}}r}
            & \multicolumn{3}{c}{IceCube}                                                    & \multicolumn{2}{c}{galactic center}                       & $e^\pm$                     & diffuse $\gamma$            & \multicolumn{2}{c}{Lab} \\[0cm]
    \cline{2-4} \cline{5-6} \cline{7-7} \cline{8-8} \cline{9-10}
            & \multicolumn{2}{c}{Boosted DM}                   & Secondary $\nu$             &                                                           &                             &                             &              & \\[0cm]
    \cline{2-3} \cline{4-4}
    $m_a$   & $m_\phi$ & $g_{Y_b}^2 g_\chi^2 f_\phi/\tau_\phi$ & $g_\chi^2 f_\phi/\tau_\phi$ & $m_\chi$ & $\ev{\sigma v_\text{rel}}_{b\bar{b}} f_\chi^2$ & $g_\chi^2 f_\phi/\tau_\phi$ & $g_\chi^2 f_\phi/\tau_\phi$ & Model        & $g_{Y_b}$ \\[0cm]
    [GeV]   & [PeV]    &  [$10^{-26}$\,s$^{-1}$]               & [$10^{-26}$\,s$^{-1}$]      & [GeV]    &     [$10^{-26}$\,cm$^3/$s]                     & [$10^{-26}$\,s$^{-1}$]      & [$10^{-26}$\,s$^{-1}$]      &              &    \\
    \hline
    \mr{12} & \mr{4.5} & \mr{0.32}                             & \mr{0.44}                   & \mr{30}  & \mr{1}                                         & \mr{$\lesssim 4.4$}         & \mr{$\lesssim 0.76$}          & \small MSSM-like    & $\lesssim 0.3$ \\
            &          &                                       &                             &          &                                                &                             &                             & \small Flipped      & $\lesssim 0.013 \tan\beta$ \\
            &          &                                       &                             &          &                                                &                             &                             & \small Vector-quark & $\lesssim 20$\\
    \hline
    \mr{80} & \mr{3.9} & \mr{2.8}                              & \mr{1.2}                    & \mr{30}  & \mr{2}                                         & \mr{$\lesssim 3.5$}         & \mr{$\lesssim 1.44$}          & \small MSSM-like    & $\lesssim 3$   \\
            &          &                                       &                             &          &                                                &                             &                             & \small Flipped      & $-$\\
            &          &                                       &                             &          &                                                &                             &                             & \small Vector-quark & $\lesssim 20$ \\
  \end{tabular}
  \end{ruledtabular}
  \caption{Summary of constraints on the boosted DM scenario for two different benchmark
    values for the mass $m_a$ of the pseudoscalar that mediates interactions between the
    light DM species $\chi$ and the SM.  Since IceCube sees both the scattering of
    highly boosted $\chi$ particles and secondary neutrinos from $\phi \to \chi \bar\chi
    + (a \to b \bar{b})$ decay, the experiment constrains two independent combinations
    of the pseudoscalar couplings to DM ($g_\chi$)
    and $b$ quarks ($g_{Y_b}$), the lifetime of the heavy DM particle, $\tau_\phi$,
    and its fractional
    abundance in  the Universe, $f_\phi$. Note that we always assume here that $a$
    couplings to SM fermions other than the $b$ quark are negligible.
    Requiring that the galactic center gamma ray excess can be explained constrains
    the light DM mass $m_\chi$ and an additional combination of coupling constants.
    Further constraints come from secondary $e^\pm$ and $\gamma$ rays from
     $\phi \to \chi \bar\chi + (a \to b \bar{b})$ and from laboratory searches for
     the pseudoscalar mediator $a$.}
  \label{tab:constraints}
\end{table}

In summary, we have discussed the possibility that the high energy event excess
observed by the IceCube collaboration is explained by the scattering of highly
boosted DM particles on atomic nuclei in the detector.  We have constructed a simple
toy model in which a DM particle $\phi$ with a mass of order PeV can decay into
a much lighter DM species $\chi$.  The $\chi$ particles, in turn, interact with
atomic nuclei through a $t$-channel mediator $a$, thus explaining the IceCube signal
at PeV energies.

The experimental constraints on this toy model are summarized in
table~\ref{tab:constraints} for two different benchmark values of the
pseudoscalar mass $m_a$.  At both benchmark points, we have assumed that the
mediator $a$ has significant coupling to $b$ quarks, while its couplings to
light quarks and to leptons are suppressed.  This is naturally realized in
UV-complete models with either an extended Higgs sector or with the
introduction of vector-like quarks (see sec.~\ref{sec:models}).  The highest
energy events in IceCube set the scale for the heavy DM mass $m_\phi$ and the
normalization of the scattering cross section. At lower energy, IceCube is
sensitive to the secondary neutrino flux from the 3-body decay $\phi \to \chi
\bar\chi + (a \to b \bar{b})$, see fig.~\ref{fig:IceCubeSpectrum}.  This
provides a constraint on the branching ratio for this decay.  Since the same
decay also leads to secondary electron/positron and gamma ray fluxes, $e^\pm$
and $\gamma$ ray data from AMS-02, Fermi-LAT, HESS and several air shower arrays
provide a constraint on its
branching ratio as well.  Moreover, the boosted DM scenario is constrained by
searches for the new pseudoscalar particle $a$ in flavor physics experiments and at high
energy colliders.

We have shown that, besides explaining a population of high energy events in
IceCube, the boosted DM scenario can simultaneously also account for the gamma
ray excess observed in Fermi-LAT data from the direction of the galactic
center.  This is possible because the light DM species $\chi$ can have a
non-negligible thermally produced relic abundance, and can annihilate in the
Milky Way today.  Fermi-LAT data then identifies a preferred range for the
light DM mass $m_\chi$ and its couplings to ordinary matter.

The boosted DM scenario shares some features with interpretations of the
IceCube data in terms of DM decay directly to SM particles, including
neutrinos.  First, the morphology of the signal is similar in the two cases,
with a mild peak expected in the galactic center region.  Moreover, it is worth mentioning
that the most recent IceCube data~\cite{Aartsen:2014muf} provides a mild
hint at a bump-like feature at $\sim 80$~TeV from the southern sky.  Since this
is where the galactic center is located, such a bump could be explained by
the secondary neutrino flux in  the boosted DM scenario. The second common
feature between boosted DM and more conventional decaying DM explanations of the
IceCube data is the rapid drop of the signal at energies larger than half of the
heavy DM mass.  With more statistics collected, these features can help to
distinguish DM interpretations of the IceCube data from an interpretation
in terms of isotropic astrophysical neutrino emission.

A unique feature of the boosted DM scenario is the prediction that, at PeV
energies, where the IceCube signal is explained by scattering of boosted DM
particles, only shower-like events should be observed. At lower energies $\sim
100$~TeV, however, where the secondary neutrino flux from the 3-body decay
$\phi \to \chi \bar\chi + (a \to b \bar{b})$ contributes, both shower and
track-like events are predicted, with a ratio very similar to the one expected
from astrophysical neutrino sources.  Between the two populations of events, a
mild dip in the energy spectrum is predicted.  These features distinguish the
boosted DM scenario from astrophysical explanations of the IceCube data and
from interpretations in terms of neutrinos from DM decay.

\section*{Acknowledgments}

It is a pleasure to thank Carlos Arg\"uelles, Sergio Palomares Ruiz,
Hubert Spiesberger and Wei Xue for useful
discussions.  We also
gratefully acknowledge discussions with Stefano Morisi, who
independently had the idea of interpreting the high-energy IceCube events as a DM
``direct detection'' signal.
JK and JL are supported by the German Research Foundation (DFG) under
Grant No.\ \mbox{KO~4820/1--1}. JK would like to thank CERN for hospitality during the
final stages of this work.

\appendix
\section{Details on the calculation of effective detector mass}
\label{app:meff}

\begin{figure*}
  \includegraphics[width=0.5\textwidth]{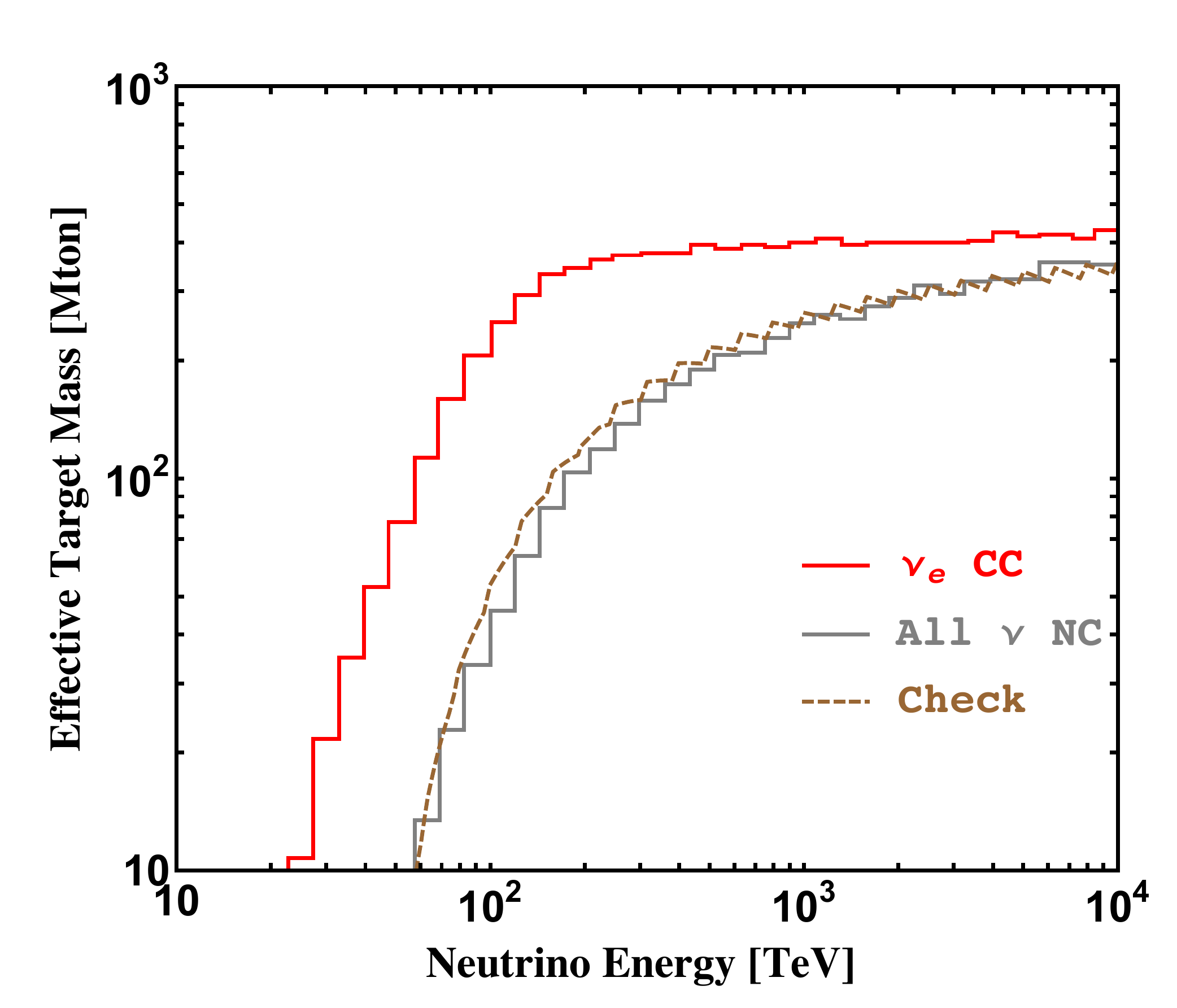}
  \caption{Effective target mass for neutrino interactions in IceCube
    as a function of the incoming neutrino energy $E_\nu$ for CC $\nu_e$ (red solid)
    and NC (gray solid) interactions~\cite{Aartsen:2013jdh}.
    The dashed brown line shows our prediction for the effective target mass
    in the NC case from eq.~\eqref{eq:MeffNCEv}, which is in excellent agreement
    with the results from~\cite{Aartsen:2013jdh}.}
  \label{fig:Meff-check}
\end{figure*}

Here, we discuss how we obtain the effective detector mass of IceCube,
$M^\text{NC/CC}(E_\nu)$ which appears in eq.~\eqref{eq:Nchi}.
We use fig.~7 from ref.~\cite{Aartsen:2013jdh}, which shows
the effective detector mass as a function of
neutrino energy rather than the deposited energy $E_\text{dep}$. However, we
can exploit the fact that in charged current (CC) interactions of electron
neutrinos, all the neutrino energy (including both the energy transferred to the
hadronic system and the energy of the produced electron) is deposited in the detector.
Electron neutrinos produce shower events very similar to neutrino or DM neutral
current scattering, hence we can assume
\begin{align}
  M^\text{NC}(E_\text{dep}) = M_{\nu_e}^\text{CC}(E_\nu) \big|_{E_\nu=E_\text{dep}} \,.
  \label{eq:MeffEdep}
\end{align}
We have verified the validity of this assumption by checking that we can use
$M^\text{NC}(E_\text{dep})$ obtained this way to reproduce the effective detector
mass for NC \emph{neutrino} interactions according to the convolution formula
\begin{align}
  M^\text{NC}(E_\nu) = \int_0^{E_\nu} \! dE_\text{dep} \,
    M^\text{NC}(E_\text{dep}) \, \frac{1}{\sigma_\nu^\text{NC}(E_\nu)}
    \, \frac{d\sigma_\nu^\text{NC}(E_\nu, E_\text{dep})}{dE_\text{dep}} \,,
  \label{eq:MeffNCEv}
\end{align}
where $d\sigma_\nu^\text{NC}(E_\nu, E_\text{dep}) / dE_\text{dep}$
is the differential cross section for NC neutrino interaction and
$\sigma_\nu^\text{NC}(E_\nu)$ is the corresponding total cross
section~\cite{Gandhi:1995tf}.
In fig.~\ref{fig:Meff-check}, we compare our result for $M^\text{NC}(E_\nu)$
with the IceCube data (fig.~7 in \cite{Aartsen:2013jdh}), and find excellent
agreement. Our results are also in agreement with the dedicated fitting
result from \cite{Palomares-Ruiz:2015mka}.

\bibliographystyle{JHEP}
\bibliography{./boosted-dm}

\end{document}